\documentclass[11pt]{article}
\usepackage{cite}
\usepackage{amssymb,amsmath,amsthm,mathrsfs}

\def\oper{{\mathchoice{\rm 1\mskip-4mu l}{\rm 1\mskip-4mu l}
{\rm 1\mskip-4.5mu l}{\rm 1\mskip-5mu l}}}
\usepackage[english]{babel}
\usepackage{graphicx}
\usepackage{bm}
\usepackage{color}
\usepackage{dsfont}
\usepackage{subfigure}
\usepackage{tikz}
\usepackage{float}
\usepackage[utf8]{inputenc}
\usepackage[T1]{fontenc}
\usepackage{amsfonts}
\usepackage{amsmath} 
\usepackage{hyperref}
\def\<{\langle}
\def\>{\rangle}
\def\bra#1{\langle #1 |}
\def\ket#1{| #1 \rangle}

\newtheorem{Theorem}{Theorem}

\newtheorem{Definition}{Definition}
\newtheorem{Remark}{Remark}

\def\<{\langle}
\def\>{\rangle}

\newcommand{\Tr}{\operatorname{Tr}}

\begin{document}

\title{\bf Fifty Years of the GKLS Master Equation: \\ Foundations and Early Developments}


\author{Fabio Benatti \\
Dipartimento di Fisica, Università di Trieste, Trieste, Italy \\
INFN, Sezione di Trieste, Trieste, Italy\\[1ex]
email: benatti@ts.infn.it \\ \\
Dariusz Chru\'sci\'nski\\
Institute of Physics, Nicolaus Copernicus University, Toru\'n, Poland
\\[1 ex]
e-mail: darch@fizyka.umk.pl
\\ \\
Saverio Pascazio \\
Dipartimento di Fisica, Universit\`{a} di Bari, Bari, Italy \\
INFN, Sezione di Bari, Bari, Italy\\[1ex]
e-mail: saverio.pascazio@uniba.it}
\author{Fabio Benatti$\hspace{0.4mm}^{1,2}$\thanks{Email: benatti@ts.infn.it}\;, \hspace{1.2mm}
\hspace{1.2mm}
Dariusz Chru\'sci\'nski$\hspace{0.4mm}^3$\thanks{Email:darch@fizyka.umk.pl}\,;\hspace{1.2mm}
\hspace{1.2mm}
Saverio Pascazio$\hspace{0.4mm}^{4,5}$\thanks{Email: saverio.pascazio@uniba.it}}

\date{
{\normalsize
$^1$\textit{Department of Physics, University of Trieste, I-34151, Trieste, Italy}\\	
\vspace{2mm}
$^2$\textit{Istituto Nazionale di Fisica Nucleare, Sezione di Trieste, I-34151,\\
Trieste, Italy}\\[1ex]e-mail:benatti@ts.infn.it}\\
\vspace{2mm}
$^3$\textit{Institute of Physics, Nicolaus Copernicus University, Toru\'n, Poland}\\[1ex]e-mail:  darch@fizyka.umk.pl\\
\vspace{2mm}
$^4$\textit{Department of Physics, University of  Bari, I-70126 Bari, Italy}\\
\vspace{2mm}
$^5$\textit{Istituto Nazionale di Fisica Nucleare, Sezione di Bari, I-70126,\\
Bari, Italy}\\[1ex]
e-mail: saverio.pascazio@uniba.it}

\maketitle

\begin{abstract}
    The year 2026 marks the fiftieth anniversary of the two seminal papers in which Gorini, Kossakowski, and Sudarshan, and independently Lindblad, characterized the generators of completely positive quantum dynamical semigroups. We revisit the conceptual and historical route leading to the Gorini–Kossakowski–Lindblad–Sudarshan (GKLS) master equation and examine several {of its} early developments
    {and} consequences. Our aim is to show how several initially separate mathematical and physical ideas converged in the GKLS structure theorem, {immediately opening the way to manifold consequences for the description of dissipative  quantum phenomena} and how that theorem, fifty years later, remains both a foundational result and an active framework for open quantum dynamics.
\end{abstract}

\tableofcontents

\section{Introduction}

The year 2026 marks the fiftieth anniversary of two papers that changed the
mathematical language of open quantum dynamics.  In May 1976, Vittorio
Gorini, Andrzej Kossakowski, and George Sudarshan published their
characterization of completely positive dynamical semigroups for finite
dimensional systems~\cite{GKS}.  One month later, G\"oran Lindblad published an
operator-algebraic characterization covering uniformly continuous quantum
dynamical semigroups on $\mathcal{B}(\mathcal{H})$, where the Hilbert space $\mathcal{H}$ could be infinite dimensional \cite{L}.  The two articles had
been submitted independently, within less than three weeks of one another,
in March and April 1975.  Their results are now jointly embodied in the
Gorini--Kossakowski--Lindblad--Sudarshan (GKLS) master equation.

In its familiar Schr\"odinger-picture form, the equation for the density operator $\rho_t$ reads $\dot{\rho}_t = \mathcal L(\rho_t)$, where the celebrated GKLS generator has the following form (following \cite{GKS,L} we set $\hbar=1$)
\begin{equation}
  \mathcal L(\rho)
  =- i[H,\rho]
  +\sum_j\left(
    V_j\rho V_j^\dagger
    -\frac12\{V_j^\dagger V_j,\rho\}
  \right).
  \label{eq:GKLS-introduction}
\end{equation}
Equivalently, $\{\exp(t\mathcal L)\}_{t\geq0}$ defines a completely positive,
trace-preserving semigroup.  Equation~\eqref{eq:GKLS-introduction} separates
the reversible Hamiltonian contribution from the irreversible part while
ensuring that the evolution remains physically meaningful not only for an
isolated system, but also when the system is coupled to an arbitrary
ancilla.  This characterization transformed the master equation
from a collection of phenomenological models into a structural theorem.

\begin{figure}[h!]
    \centering
    \includegraphics[width=11cm]{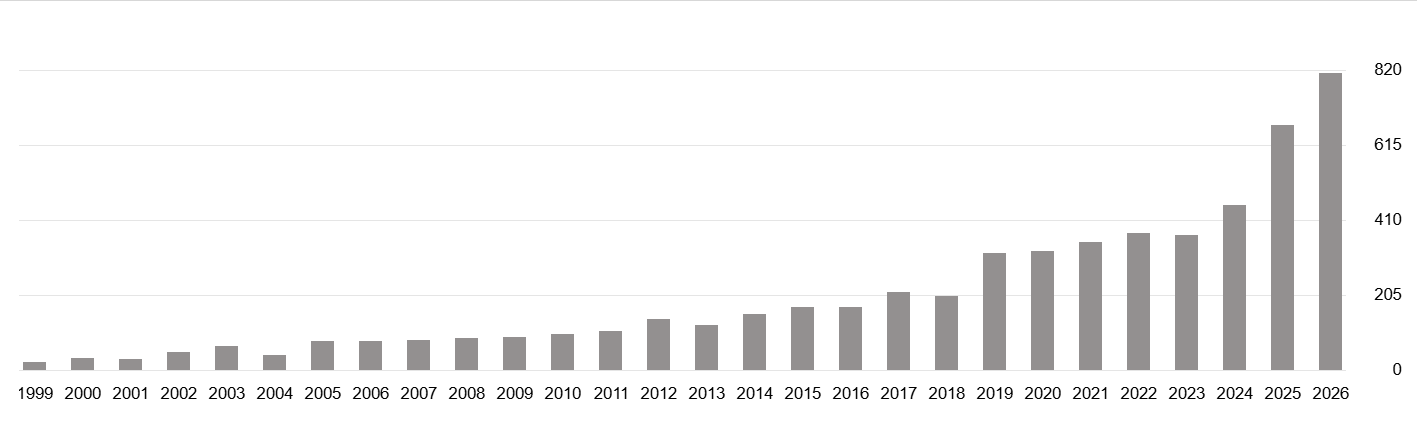} \includegraphics[width=11cm]{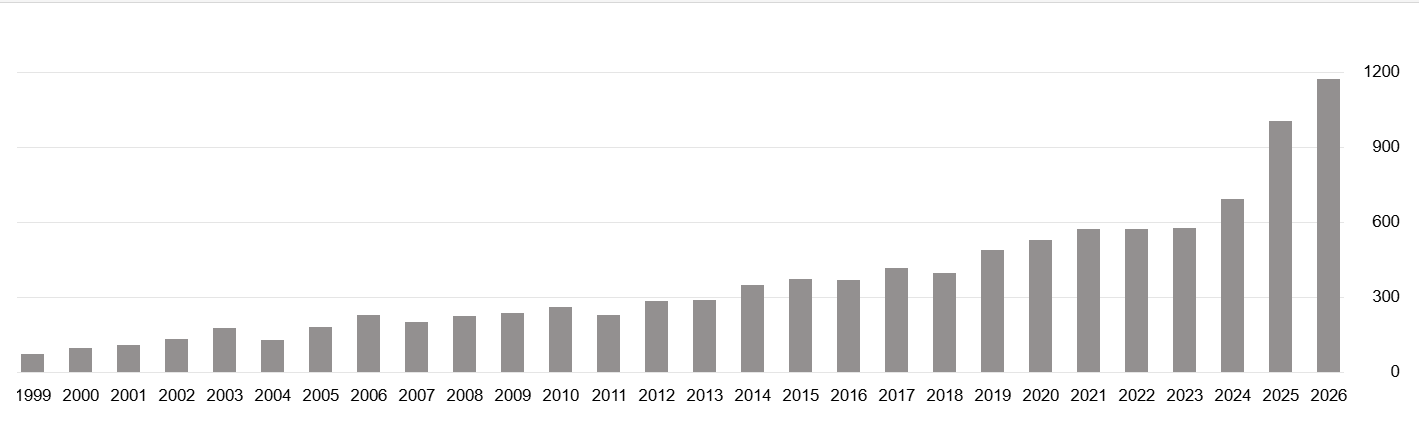} 
    \caption{Citations counts (Scholar Google 20th August, 2026): upper panel -- GKS paper; lower panel -- Lindblad paper. }
   \label{Fig-1}
\end{figure}

Half a century later, the GKLS equation remains the standard local-in-time
description of Markovian open quantum systems.  It underlies quantum optics,
atomic and condensed-matter physics, nonequilibrium statistical mechanics,
quantum information theory, quantum thermodynamics, dissipative state
engineering, error correction, and the modelling of present-day quantum
devices.  Its role has in fact grown rather than diminished.  The annual
citation counts displayed in Figure~\ref{Fig-1} provide a simple quantitative
illustration: both original papers have attracted sharply increasing
attention during the last decade, with the largest yearly numbers reached
around the anniversary itself.  Citation counts are of course not a measure of conceptual depth, but
here they vividly show that the 1976 structural theorem is not merely a historical
landmark; it remains an active working tool.

The now-canonical compact formula (\ref{eq:GKLS-introduction}) can obscure the long and intricate route by which it was reached. The GKLS Theorem  brought together
several developments: the density operator as the appropriate description of statistical quantum states;
linear maps preserving density matrices; the notions of positivity and complete positivity; the concept of one-parameter semigroups and their generators; and microscopic limiting procedures capable of
producing an irreversible dynamics from an underlying reversible theory. These ideas were developed
in different mathematical and physical communities, often with different terminology and with only
partial awareness of one another. 

{The fortieth anniversary of the GKLS equation was marked by two
special issues of \emph{Open Systems \& Information Dynamics}
\cite{OSID40GKLS}. These included our historical account \emph{A Brief History of the GKLS Equation}
\cite{40GKLS} and the retrospective contribution
of Accardi and Lu \emph{The First 40 Years of {GKSL} Generators and Some Proposal for the Future} \cite{AccardiLu2017}. } The
present contribution returns to that history on the occasion of the
fiftieth anniversary, but with a broader purpose.  We look further back at
results that anticipated essential pieces of the 1976 structure theorems, and we
follow several lines of development that began immediately after the
theorems were published.  

The prehistory begins with the 1961 analysis of Sudarshan, Mathews, and Rau,
who represented the evolution of density matrices by a linear dynamical
matrix and formulated positivity and trace preservation at the level of
that matrix~\cite{SMR61}.  Related work by Jordan and Sudarshan further developed the
theory of dynamical mappings~\cite{Jordan-Sudarshan-61}.  Bausch's 1966 investigation of evolution
equations for non-isolated systems~\cite{Bausch1966}, and the 1969 analysis by Belavin,
Zel'dovich, Perelomov, and Popov~\cite{Belavin1969}, contained remarkably early forms of
Markovian relaxation equations.  Kossakowski's papers of 1972--1973 then
placed non-Hamiltonian evolution within a systematic semigroup framework,
including general generator conditions and a detailed analysis of the
two-level system~\cite{AK-1,AK-2,AK-1972}; related work with Ingarden further connected this
framework with nonequilibrium statistical mechanics~\cite{I-K}.  In parallel,
Davies' weak-coupling program provided a microscopic route from
system--reservoir dynamics to a quantum Markov semigroup~\cite{Davies1974}.  These results
show that the GKLS theorem had a rich prehistory, even though the {fundamental} 
role of complete positivity had not yet been universally recognized.

\begin{figure}[h!]
    \centering
    \includegraphics[width=7cm]{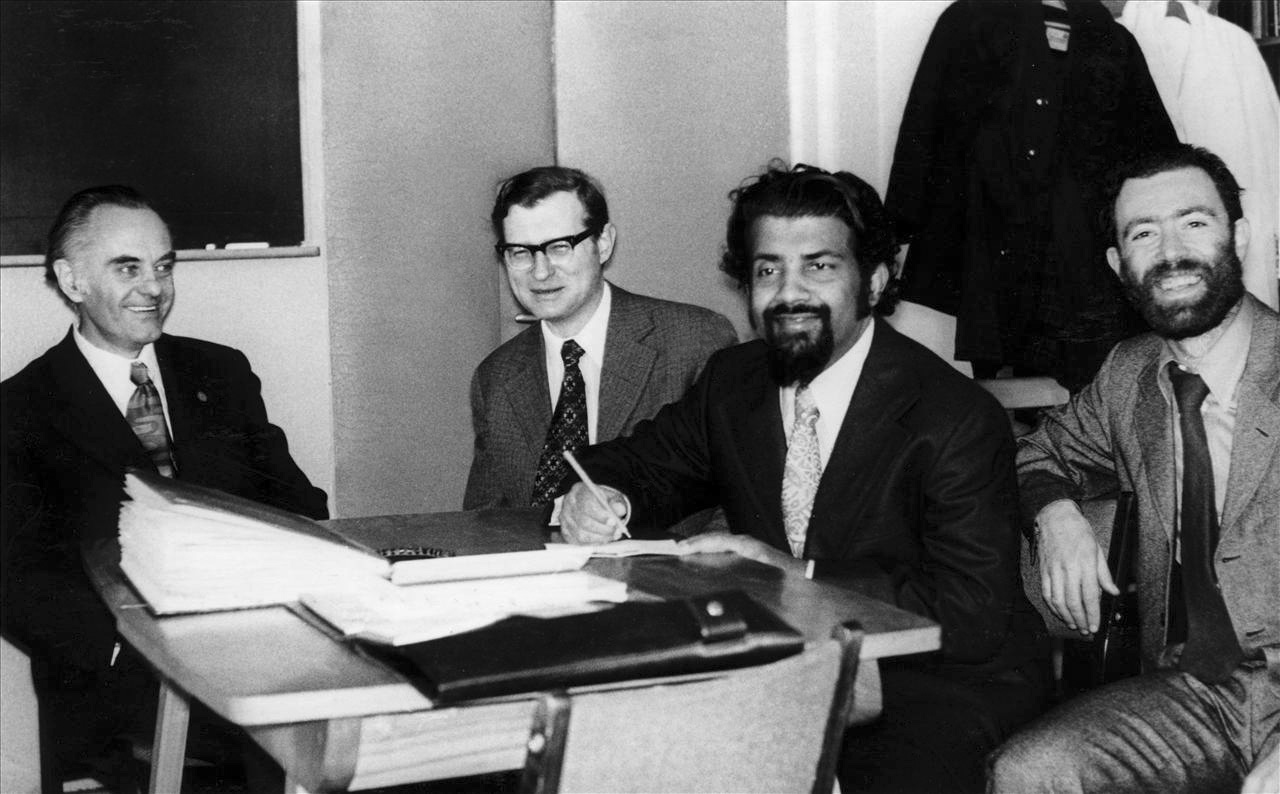} \includegraphics[width=6.1cm]{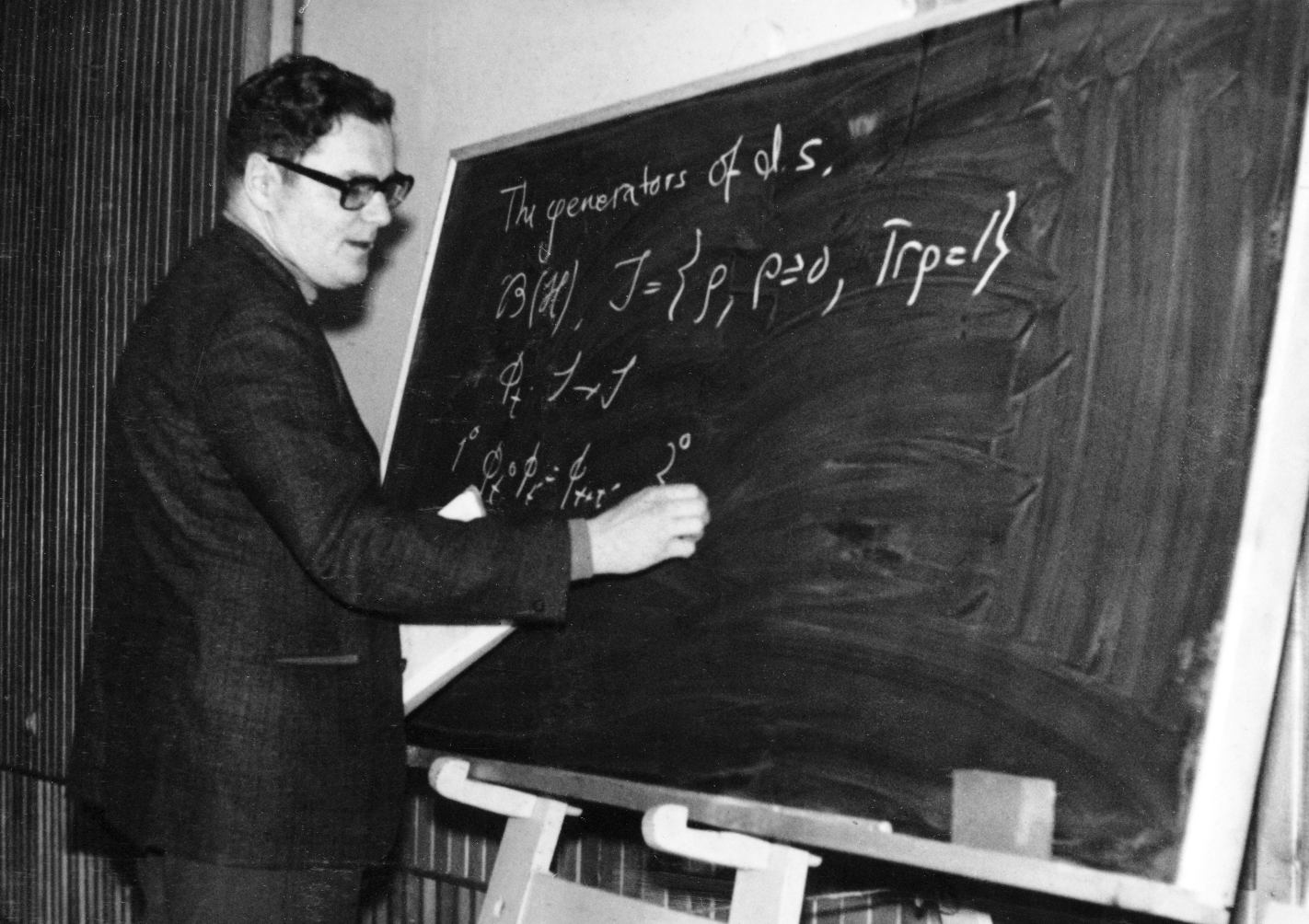} 
    \caption{Left panel: picture taken in Prof. Ingarden’s office (December 1975). From left
to right: Roman Ingarden, Andrzej Kossakowski, George Sudarshan and
Vittorio Gorini. Right panel:  picture taken during the Symposium on Mathematical Physics in Toru\'n (December 1974) -- G\"oran Lindblad lecturing about quantum dynamical semigroups. }
   \label{Fig-2}
\end{figure}

The decisive convergence took place during 1973--1975.  Gorini encountered
the theory of positive and completely positive maps at the 1973 Marburg
meeting and subsequently discussed it with Kossakowski and Sudarshan in
Austin.  Lindblad approached the problem independently through the theory of
operator algebras and dissipative semigroups.  Personal encounters were
nonetheless part of the story: Lindblad lectured on quantum dynamical
semigroups at the Symposium on Mathematical Physics in Toru\'n in December
1974, and Gorini visited him in Stockholm in January 1975 on his return from
Texas.  The photographs collected in Figure~\ref{Fig-2} give this abstract history a
human scale.  The right panel shows Lindblad lecturing in Toru\'n in 1974;
the left panel, taken in Roman Ingarden's office in December 1975, shows
Ingarden, Kossakowski, Sudarshan, and Gorini together.  By then the two
seminal manuscripts had already been submitted. 

The GKS and Lindblad theorems \cite{GKS,L} solved closely related problems by genuinely
different routes.  GKS worked in the Schr\"odinger picture and in finite
dimension.  Expanding the generator in a traceless operator basis, they
identified positivity of the coefficient matrix---now called the
Kossakowski matrix---as the necessary and sufficient condition for complete
positivity.  Lindblad worked primarily in the Heisenberg picture and used
the infinitesimal dissipativity condition associated with a Kadison-Schwarz inequality. 
 The equivalence of these formulations
is now routine, but their derivations reveal two complementary ways of
thinking: one in terms of a positive quadratic form of coefficients (Kossakowski matrix) \cite{GKS}, the
other in terms of the intrinsic geometry of operator algebras \cite{L}.

The publication of the GKLS structure theorem was a beginning rather than
an endpoint.  
In the same year Alicki formulated quantum detailed balance
for non-Hamiltonian systems~\cite{Alicki_1976}.  Spohn and Frigerio established fundamental
criteria for stationary states, irreducibility, relaxation, and approach to
equilibrium~\cite{Spohn_1976,Spohn_1977, Frigerio_1977, Frigerio_1978}.  Lindblad's later work connected quantum
regression with the Markov property~\cite{Lindblad1979}, while the Hudson--Parthasarathy
calculus supplied a stochastic dilation of quantum Markov dynamics~\cite{HudsonParthasarathy1984}.
The equation also entered discussions of fundamental irreversibility and
information loss, notably in the analysis of Banks, Susskind, and
Peskin~\cite{Banks1984}.  Finally, the Pechukas--Alicki debate on initially correlated
system--environment states clarified both the power and the proper domain of
the complete-positivity assumption~{\cite{Pechukas,AvsP95,Pechukas94}} 

As stressed in Ref.~\cite{40GKLS}, equations that look similar in modern notation may
have arisen from different hypotheses, mathematical tools, and physical
questions.  Conversely, results expressed in unfamiliar language may
contain an essential part of what later became the standard framework.  We
therefore combine historical reconstruction with enough technical detail to
make these connections explicit, retaining original formulations whenever
they illuminate how the concepts evolved.

The paper is organized into three main historical parts. Section~2,
devoted to the period preceding the GKLS theorem, traces the development
from the matrix representation of dynamical maps introduced by Sudarshan,
Mathews, and Rau, through the early master-equation and semigroup analyses
of Bausch, Belavin and co-workers, and Kossakowski, to Davies' rigorous
weak-coupling derivation of a quantum dynamical semigroup. Section~3
examines the two landmark structure theorems of 1976, emphasizing the
finite-dimensional formulation of Gorini, Kossakowski, and
Sudarshan and the operator-algebraic approach of Lindblad. Section~4
follows the early developments initiated by these results: Alicki's
formulation of quantum detailed balance; the Spohn--Frigerio theory of
relaxation and approach to equilibrium; Spohn's analysis of entropy
production and irreversible quantum thermodynamics; Lindblad's connection
between quantum regression and Markovianity; the Hudson--Parthasarathy
quantum stochastic calculus; the analyses of fundamental information loss
by Ellis--Hagelin--Nanopoulos--Srednicki and
Banks--Susskind--Peskin; the quantum-jump unravelling of
Dalibard--Castin--M{\o}lmer; and the Pechukas--Alicki debate concerning
complete positivity in the presence of initial system--environment
correlations. Section~5 concludes with a perspective on the continuing
legacy of the GKLS equation.



\section{Before GKLS: towards open quantum dynamics}

\subsection{Sudarshan's matrix representations (1961)}
\label{S:Sudarshan}

The modern theory of quantum dynamical maps can be traced back to the remarkable 1961 paper by Sudarshan, Mathews and Rau entitled \emph{Stochastic Dynamics of Quantum-Mechanical Systems} \cite{SMR61}. The authors posed a question that was far ahead of its time: what is the most general dynamical law describing the time evolution of a quantum state? In the abstract
the authors state:

"{\em The most general dynamical law for a quantum mechanical system
with a finite number of levels is formulated. A fundamental role
is played by the so-called ``dynamical matrix'' whose properties
are stated in a sequence of theorems. A necessary and sufficient
criterion for distinguishing dynamical matrices corresponding to
a Hamiltonian time-dependence is formulated. }"

Their starting point was the observation that the density matrix, rather than the Schr\"odinger wave function, provides the most general description of a quantum state. Instead of assuming unitary time evolution from the outset, they searched for the most general linear mapping

\begin{equation} \label{PHI}
    \rho(t) = \Phi(t,t_0)[\rho(t_0)] , \quad t \geq t_0 ,
\end{equation}
that transforms density operators into density operators. Linearity was motivated by the requirement that incoherent mixtures of states remain incoherent mixtures under time evolution.
Fixing an orthonormal basis $\{|r\rangle\}_{r=1}^n$, the map is represented by
an $n^2\times n^2$ matrix $A =[A_{rs,r's'}]$ through

\begin{equation}\label{}
  \rho_{rs}(t) = \sum_{r',s'} A_{rs,r's'}(t,t_0) \rho_{r's'}(t_0) ,
\end{equation}
where the matrix $A_{rs,r's'}(t,t_0)$ is defined by

\begin{equation}
    A_{rs,r's'}(t,t_0) = \langle r| \Phi(t,t_0)[|r'\rangle \langle s'|] | s\rangle .
\end{equation}
{Set $\Phi:=\Phi(t,t_0)$ for sake of simplicity; t}he physical requirements on $\Phi$ are

\begin{eqnarray}
    && \Phi[X]^\dagger = \Phi[X^\dagger] \ , \ \ \ \ \mbox{ (Hermiticity)} \nonumber \\ 
    &&  \langle y |\Phi[|x\rangle \langle x|] | y\rangle \geq 0 \ , \ \ \ \ \mbox{(positivity)} \\ 
    && {\rm Tr}\, \Phi[X] = {\rm Tr}\, X \ , \ \ \ \ \mbox{(trace-preservation)}\ , \nonumber 
\end{eqnarray}
In the $A$-representation they become, respectively,

\begin{eqnarray}\label{A}
 && A_{sr,s'r'} = {A^*_{rs,r's'}} \ , \ \ \ \ \mbox{ (Hermiticity)} \label{A1} \\
 && \sum_{r,s,r',s'}{{x}^*_{s'}\, x_{r'} \, A_{rs,r's'}\, y^*_{r} \, {y_{s}}} \geq 0 \ , \ \ \ \ \mbox{(positivity)} \label{A2}\\
 \label{A3} &&  \sum_r A_{rr,r's'} = \delta_{r's'} \ , \ \ \ \ \mbox{(trace-preservation)}\ . 
\end{eqnarray}
Thus $A$ is simply the matrix of the superoperator in Liouville space defined by the vectorization procedure:

\begin{equation}
    \widehat{\Phi} |\rho\rangle \! \rangle = |\Phi(\rho) \rangle \! \rangle ,
\end{equation}
that is, $ A_{sr,s'r'} = \langle s \otimes r| \widehat{\Phi}  | s' \otimes r' \rangle$. 
If one knows how to represent the superoperator $\widehat{\Phi}$, or equivalently, its $A$-matrix satisfying the above properties, then one finds the dynamical law for the system's density operator. However, as {the} authors remarked ``{\it to display these properties in a more transparent fashion, as well as for further development, it is
advantageous to introduce another matrix $B$ related to $A$ and defined by}''

\begin{equation}\label{realignment}
  B_{rr',ss'}  := A_{rs,r's'}  .
\end{equation}
Let us observe that the matrix $B$ is nothing but the Choi matrix \cite{Choi75} of $\Phi$ and it was $B$ that the authors called the \emph{dynamical matrix}.  Now, the authors make the surprising statement ``{\it it immediately follows that $B$ is Hermitian and positive
semidefinite}'', that is, conditions (\ref{A1}) and (\ref{A2}) transform to 

\begin{eqnarray}\label{B}
 && B_{rr',ss'} = {B^*_{ss',rr'}} \ , \ \ \ \ \mbox{ (Hermiticity)}  \label{B1} \\
 && \sum_{r,s,r',s'} z^*_{rr'} B_{rr',ss'} z_{ss'} \geq 0  \ , \ \ \ \ \mbox{(positivity)} \label{B2} .
\end{eqnarray}
Note, however, that condition (\ref{A2}) is not equivalent to (\ref{B2}); {indeed, 
not all $z_{ss'}$ split as $z_{ss'}=y_s\,x^*_{s'}$}. Instead, it is equivalent to 

\begin{equation}
    \sum_{r,s,r',s'} { y^*_{r} x_{r'}B_{rr',ss'}  y_{s}x^*_{s'}} \geq 0 .   \label{B2a}
\end{equation}
Condition (\ref{B2a}) is much weaker than (\ref{B2}) and it implies only block-positivity of the $B$-matrix which is equivalent to positivity of the original map $\Phi$. Positive semidefiniteness of $B$
is strictly stronger and, in modern terminology, is equivalent to complete
positivity of $\Phi$ \cite{Stormer1, Stormer1974, Paulsen2002, Stormer2013}.  The distinction is already visible for transposition:
the map $X\mapsto X^{\mathsf T}$ is positive and trace preserving, whereas its
dynamical matrix is the flip operator, which has negative eigenvalues (cf. also interesting historical remarks by Kavan Modi \cite{Modi_2019}).

If one adopts the stronger condition (\ref{B2}), the spectral
resolution of the dynamical matrix takes the form

\begin{equation}
 B=\sum_\alpha\mu_\alpha
 |W_\alpha\rangle\!\rangle\langle\!\langle W_\alpha|,
 \qquad \mu_\alpha\geq0,
 \label{eq:smr-B-spectral}
\end{equation}
or, in the component notation used in the original paper,

\begin{equation}
 B_{rr',ss'}=\sum_\alpha\mu_\alpha
 (W_\alpha)_{rr'}(W_\alpha)_{ss'}^*.
 \label{eq:smr-B-components}
\end{equation}
Upon undoing the realignment {in~\eqref{realignment}}, this is precisely the operator-sum
representation

\begin{equation}
 \Phi(X)=\sum_\alpha\mu_\alpha W_\alpha XW_\alpha^\dagger,
 \qquad
 \sum_\alpha\mu_\alpha W_\alpha^\dagger W_\alpha=\oper .
 \label{eq:smr-operator-sum}
\end{equation}
Thus the canonical form derived in the 1961 paper is correct for completely
positive maps, even though the authors presented it as a consequence of mere
positivity.  For a positive map which is not completely positive, $B$ is still
Hermitian and therefore has a spectral decomposition, but some eigenvalues
$\mu_\alpha$ are negative; Eq.~\eqref{eq:smr-operator-sum} then becomes a
signed operator-sum representation and no longer proves positivity term by
term.

The paper also identified the unitary subclass.  If
$\Phi(X)=UXU^\dagger$, then

\[
 B=|U\rangle\!\rangle\langle\!\langle U|,
 \qquad \operatorname{Tr}B=n,
 \qquad B^2=nB.
\]
Conversely, a dynamical matrix satisfying the partial-trace condition and
$B^2=nB$ has one nonzero eigenvalue, equal to $n$, and hence is of the above
rank-one form; the partial trace {condition~\eqref{A3}} then forces $U$ to be unitary.  Therefore a
Hamiltonian evolution is characterized by a rank-one dynamical matrix, whereas
non-Hamiltonian stochastic dynamics requires higher dynamical rank.  This was
one of the principal theorems of the 1961 paper.  The authors further described
the convexity of dynamical matrices and discussed random-unitary mixtures,
reset maps, and longitudinal and transverse spin relaxation.

\begin{Remark}
Equation~\eqref{eq:smr-operator-sum} appeared more than a decade before the
corresponding matrix argument of Choi~\cite{Choi75} and before Kraus's systematic use of
operator-sum representations~\cite{Kraus1971, Kraus1983}.  Stinespring's dilation theorem \cite{Stinespring1955} had already
been proved in 1955, but the finite-dimensional dynamical-matrix derivation of
Sudarshan, Mathews and Rau was independent in language and motivation.  The
historical attribution should therefore distinguish two statements: the 1961
paper anticipated the Kraus--Choi representation for the subclass $B\geq0$,
but it did not distinguish complete positivity from positivity and hence did
not establish such a representation for every positive dynamical map.  In
this light it is striking that the GKS paper \cite{GKS} did not cite \cite{SMR61}.  In \cite{40GKLS} we
proposed the name {Kraus--Stinespring--Sudarshan representation} {for the ubiquitous expansion
\begin{equation}
\label{KSSrep}
\Psi(X)=\sum_jV_j^\dag\,X\,V_j
\end{equation}
of a completely positive map $\Psi$.}
\end{Remark}

\begin{Remark} The correspondence $\Phi \to B_{rr',ss'}$ is called the Choi-Jamio{\l}kowski isomorhism. 
Actually, Jamio{\l}kowski \cite{Jamiolkowski_1972} following de Pillis \cite{dePillis} used the correspondence $\Phi \to B_{sr',rs'}$ which is related to the previous one via partial transposition. Note, however, that a map $\Phi$ is positive if and only if $\Phi \circ {\rm T}$ is also positive.  Hence, for characterizing block-positivity both correspondence are equivalent (cf. \cite{Watrous2018, BengtssonZyczkowski2017} for the detailed exposition).  
\end{Remark}

\subsection{Bausch derivation of the master equation for positive semigroup (1966)}   \label{S:Bausch}

In another remarkable but largely forgotten paper \cite{Bausch1966}, Bausch addressed a question that would later become central to the theory of open quantum systems: how can one derive a differential equation for the density matrix starting from the most general dynamical transformation of quantum states?
Actually, Bausch gave what appears to be the first systematic derivation of a
time-local quantum master equation from dynamical maps. 
{His starting point} was {what} just discussed in Section~\ref{S:Sudarshan}, 
{namely} the representation introduced by Sudarshan, Mathews and Rau, but he corrected an important aspect of their discussion: positivity of a map
implies only block positivity of its dynamical (Choi) $B$-matrix, and not positivity of
that matrix \cite{Jamiolkowski_1972}.   Consequently, the coefficients in its spectral decomposition
need only be real and may have either sign.  Bausch used the letters $A$ and
$B$ in the opposite order to Sudarshan \cite{SMR61}; here we retain the convention of
Section~\ref{S:Sudarshan}.

Let $\{A(t)\}_{t\geq0}$ be a differentiable family of Sudarshan's $A$-matrices corresponding to a  family of positive trace-preserving maps $\Phi_t : M_n(\mathbb C) \to M_n(\mathbb C)$.  Bausch's ``Markov property'' is precisely time homogeneity together with the semigroup law

\begin{equation}
\label{time-hom}
    A(t)A(\tau) = A(t+\tau) \ , \quad A(0) = \oper_n \otimes \oper_n .
\end{equation}
Equivalently, 

\[   \Phi_t \circ \Phi_s = \Phi_{t+s} \ , \quad \Phi_0 = {\rm id} .  \]
Since the dynamical matrix $B(t)$ is Hermitian, it can be diagonalized.  After
reshaping its eigenvectors into Hilbert--Schmidt orthogonal matrices
$W_\alpha(t)$, the action of the map $\Phi_t$ reads

\begin{equation}
 \rho(t+s)=\Phi_s[\rho(t)]
 =\sum_\alpha a_\alpha(s)W_\alpha(s)\rho(t)W_\alpha(s)^\dagger,
 \qquad a_\alpha(s)\in\mathbb R .
 \label{eq:bausch-dynamical-map}
\end{equation}
For a merely positive  map the numbers $a_\alpha(s)$ are not necessarily positive. 
Block positivity of $B(s)$ does not imply $a_\alpha(s)\geq0$.  Trace
preservation is equivalent to
$\sum_\alpha a_\alpha(s)W_\alpha(s)^\dagger W_\alpha(s)=\oper$.
At $s=0$ the representation may be chosen so that

\begin{equation}
 a_0(0)=1,\qquad W_0(0)=\oper ,\qquad
 a_\alpha(0)=0\quad(\alpha\ne0).
 \label{eq:bausch-initial-data}
\end{equation}
Differentiating Eq.~\eqref{eq:bausch-dynamical-map} at $s=0$, as justified by
the semigroup law, gives $\dot\rho(t)=\mathcal L[\rho(t)]$ with

\begin{equation}
 \mathcal L(\rho)=\dot W_0(0)\rho+\rho\dot W_0(0)^\dagger
 +\sum_\alpha c_\alpha W_\alpha\rho W_\alpha^\dagger,
 \qquad
 c_\alpha=\dot a_\alpha(0)\in\mathbb R, \quad
 W_\alpha=W_\alpha(0).
 \label{eq:bausch-raw-generator}
\end{equation}
Introduce the Hermitian operators

\begin{equation}
 H:=\frac{i}{2}\bigl(\dot W_0(0)-\dot W_0(0)^\dagger\bigr),
 \qquad
 K:=\frac12\bigl(\dot W_0(0)+\dot W_0(0)^\dagger\bigr).
 \label{eq:bausch-HK}
\end{equation}
The derivative of the trace-preservation identity yields
$2K+\sum_\alpha c_\alpha W_\alpha^\dagger W_\alpha=0$.  Substitution in
Eq.~\eqref{eq:bausch-raw-generator} produces Bausch's central result,

\begin{equation}
 \mathcal L(\rho)=-i[H,\rho]
 +\sum_\alpha c_\alpha
 \left(W_\alpha\rho W_\alpha^\dagger
 -\frac12\{W_\alpha^\dagger W_\alpha,\rho\}\right),
 \quad c_\alpha\in\mathbb R .
 \label{eq:bausch-generator}
\end{equation}

\begin{Remark}
Equation~\eqref{eq:bausch-generator} is an algebraic representation of the
generator, not by itself a characterization of generators of positive
semigroups.  Positivity remains encoded in the block positivity of $B(t)$ for
every $t\geq0$, and Bausch did not convert this global requirement into a
necessary and sufficient local condition on the real numbers $c_\alpha$ and
the operators $W_\alpha$.  This distinction is essential: the coefficients
$c_\alpha$ may be negative, but arbitrary negative coefficients do not define
a positive evolution. This issue was resolved few years later by Kossakowski \cite{AK-1,AK-2}, see Section~\ref{S:Kossakowski} 
\end{Remark}

\begin{Remark}
If complete positivity is imposed, then $B(t)\geq0$ and hence
$a_\alpha(t)\geq0$.  Since $a_\alpha(0)=0$ for $\alpha\ne0$, the right
derivatives satisfy $c_\alpha\geq0$.  Equation~\eqref{eq:bausch-generator}
then has exactly the GKLS form.  Thus Bausch had already found the correct
infinitesimal architecture ten years before the GKLS papers, but for positive
rather than completely positive dynamical maps; he did not prove the later
complete-positivity structure theorem.  The differentiation argument is also
essentially the one used independently by Alicki in \cite{Alicki-Lendi}. 
\end{Remark}
Bausch illustrated the point with the relaxation of a spin-$\tfrac12$ in a
magnetic field.  In modern notation his model may be written

\begin{equation}
 \mathcal L(\rho)=-i[H,\rho]
 +\gamma\bigl(\rho_0\operatorname{Tr}\rho-\rho\bigr)
 +\gamma_z\bigl(\sigma_z\rho\sigma_z-\rho\bigr),
 \qquad H=\frac{\omega}{2}\sigma_z,
 \label{eq:bausch-spin-generator}
\end{equation}
where $\rho_0=e^{-\beta H}/\operatorname{Tr}(e^{-\beta H})$ is a stationary  Gibbs state.  
The generator of the Bausch master equation~\eqref{eq:bausch-spin-generator} can be recast in the form~\eqref{eq:bausch-generator}

\[
\mathcal L(\rho)=-i[H,\rho]
+\frac{\gamma(1+M_0)}2\mathcal D_{\sigma_+}(\rho)
+\frac{\gamma(1-M_0)}2\mathcal D_{\sigma_-}(\rho)
+\left(\frac\gamma4+\gamma_z\right)\mathcal D_{\sigma_z}(\rho) ,
\]
where $\mathcal{D}_V(\rho) = V\rho V^\dagger - \frac 12 \{V^\dagger V,\rho\}$, and $M_0 = {\rm Tr}(\rho_0 \sigma_z)$. On the other hand, introducing $M_k = {\rm Tr}(\sigma_k \rho)$, {$k=x,y,z$}, one finds the following Bloch equations for {$\bm{M}=(M_x,M_y,M_z)$}:

\begin{eqnarray}
    \dot{M}_x &=& -\omega M_y - \frac{1}{T_T} M_x \ , \\
    \dot{M}_y &=& \omega M_x - \frac{1}{T_T} M_y \ , \\
    \dot{M}_z &=& - \frac{1}{T_L} (M_z - M_0) \ , 
\end{eqnarray}
where $T_L$ and $T_T$ are longitudinal and transversal relaxation times defined via 

\[    \frac{1}{T_L} = {\gamma} \ ,\ \ \frac{1}{T_T} = {\gamma} + 2 \gamma_z .\]
Positivity requires $\gamma \geq 0$ and $\gamma+ 2 \gamma_z \geq 0$, whereas complete positivity requires 
$\gamma \geq 0$ together with $\gamma+  4\gamma_z \geq 0$.  If the evolution is completely positive {the} relaxation times satisfy the following famous constraint: 
\begin{equation}
\label{hierarchy}
    T_T \leq 2 T_L \ .
\end{equation} 
Bausch's example is historically important
precisely because it displays, within a concrete Bloch equation, the gap
between positivity and the stronger condition that would later become central
to the GKLS theory.\footnote{ After the Toru\'n 2026 conference Daniel Burgarth and Hajo Leschke contacted professor Richard Bausch and sent him the slides of the presentation which one of us (DC) presented during the conference. The paper \cite{Bausch1966} contains the material of his PhD dissertation.}

\begin{Remark} Interestingly, the relation between relaxation times for completely positive semigroup (\ref{hierarchy}) was recently generalized to arbitrary $d$-level systems: denote by  $T_k$ ($k=1,2,\ldots,d^2-1)$ relaxation times and by $\Gamma_k = 1/T_k$ the corresponding relaxation rates. Then
one proves \cite{MuratoreGinanneschi2025, Chruscinski2025, MuratoreGinanneschi2026} the following universal constraint 

\begin{equation}
    \Gamma_k \leq \frac 1d \Big( \Gamma_1 + \ldots + \Gamma_{d^2-1} \Big) .
\end{equation}
For the considered qubit system due to the additional axial symmetry $T_1=T_2=T_T$ and $T_3 = T_L$. Hence (\ref{hierarchy}) implies the universal qubit relation $\Gamma_k \leq \frac 12 (\Gamma_1+\Gamma_2+\Gamma_3)$ \cite{Kimura2002}.
    
\end{Remark}

\subsection{Belavin–Zel'dovich–Perelomov–Popov analysis (1969)}
\label{S:Belavin}

In 1969 Belavin, Zel'dovich, Perelomov and Popov \cite{Belavin1969} investigated the relaxation of quantum systems possessing equidistant energy spectra, such as the harmonic oscillator, a spin in a magnetic field and the Dicke model of collectively radiating two-level atoms. Their starting point was the kinetic equation \begin{equation} \dot\rho_t = -i[V,\rho_t] +\frac{\gamma}{2}\Big[(\bar n+1) \bigl(2A\rho_t A^\dagger-A^\dagger A\rho-\rho_t A^\dagger A\bigr) +\bar n \bigl(2A^\dagger\rho_t A-AA^\dagger\rho_t - \rho_t AA^\dagger\bigr) \Big], \label{BZPP} 
\end{equation} 
where $\bar n=(e^{\beta\hbar\omega_0}-1)^{-1}$ is the thermal occupation number of the reservoir mode resonant with the system. Today Eq.~(\ref{BZPP}) is immediately recognized as a special instance of the GKLS equation with a single jump operator $A$, but in 1969 neither the concept of complete positivity nor the general structure theorem for quantum dynamical semigroups had yet been established (the authors do not cite Bausch 1966 paper \cite{Bausch1966}). The central question addressed in \cite{Belavin1969} was whether a kinetic equation of the form (\ref{BZPP}) preserves the positivity of the density matrix. 

\begin{Theorem}[Belavin et al., 1969]
If $\gamma\geq0$ and $\nu\geq0$, the evolution generated by
Eq.~\eqref{BZPP} preserves normalization, Hermiticity, and
positivity of the density matrix.
\end{Theorem}
While trace preservation and Hermiticity are obvious, positivity is much more subtle. The authors introduced the quantity $R_\psi(t)=\langle\psi|\rho(t)|\psi\rangle$,  and considered 
 
 \begin{equation} R(t)=\min_{\|\psi\|=1} R_\psi(t) , 
 \end{equation} 
 so that positivity of $\rho(t)$ is equivalent to $R(t)\ge0$. Assume that positivity holds at $t=0$ and let $|\psi\rangle$ be a vector such that $R_\psi(0)=0$, i.e. a vector lying on the boundary of the positive cone. Since positivity might only be lost through such boundary vectors, they investigated the first nonvanishing time derivative of $R_\psi(t)$. Using the structure of Eq.~(\ref{BZPP}), they proved that whenever $\langle\psi|\rho(0)|\psi\rangle=0$, the derivative $R_\psi'(0)$ is nonnegative. If $R_\psi'(0)=0$, then additional relations imply that the second derivative is also nonnegative. Consequently, 
 $ R_\psi(t)\ge0$  for sufficiently small times. By repeating the argument iteratively, positivity is preserved for all times. This geometric reasoning is remarkably close in spirit to later characterizations of generators of positive semigroups due to Kossakowski \cite{AK-1,AK-2} (cf. Section \ref{S:Kossakowski}). The authors further observed that the same proof applies to the more general equation 
 
 \begin{equation} \dot\rho_t = \sum_{i,j=1}^N \gamma_{ij} \Big( A_i\rho_t A_j^\dagger -\frac12\{A_j^\dagger A_i,\rho_t\} \Big), \label{BZPPgen} \end{equation} 
  provided the coefficient matrix $\gamma =[\gamma_{ij}]$ is positive semidefinite. Clearly, positivity of $[\gamma_{ij}]$ is sufficient but not necessary for positivity of $\rho_t$ for all $t \geq 0$.  In modern language, Eq.~(\ref{BZPPgen}) is precisely a GKLS generator written in terms of a Kossakowski matrix. Thus, seven years before the GKS and Lindblad structure theorems, Belavin \emph{et al.} had already identified positivity of the coefficient matrix as the key condition guaranteeing positivity of the dynamical evolution. 
  From a historical perspective, the work of Belavin, Zel'dovich, Perelomov and Popov may therefore be viewed as one of the closest precursors of the Gorini-Kossakowski-Sudarshan and Lindblad theorems.


\subsection{Kossakowski analysis of contractive semigroups (1972-1973)}   \label{S:Kossakowski} 

Kossakowski formalized the problem as follows: let $\mathcal{B}_1(\mathcal{H})$ denotes a Banach space of trace-class operators and let $\mathcal{B}^+_1(\mathcal{H})$ be a convex subset of positive semidefinite trace-class operators. He started with the  following

\begin{Definition} A family  $\{\Phi_t\}_{t\geq 0}$ of linear maps on $\mathcal{B}_1(\mathcal{H})$ is said to be a dynamical semi-group of a quantum system provided the following conditions are satisfied

\begin{enumerate}
    \item $\Phi_t : \mathcal{B}^+_1(\mathcal{H}) \to \mathcal{B}^+_1(\mathcal{H})$ for $t \geq 0$,

    \item $\| \Phi_t(\rho)\|_1 = \|\rho\|_1$ for all $\rho \in \mathcal{B}^+_1(\mathcal{H})$ and $t \geq 0$,

    \item $\Phi_t \Phi_s = \Phi_{t+s}$,

    \item $s\!\!-\!\!\lim_{t \to 0+} \Phi_t = {\rm id}$.
\end{enumerate}
    
\end{Definition}
Kossakowski observed \cite{AK-1} that a linear map is positive and trace-preserving if and only if it is hermiticity- and trace-preserving, and contractive w.r.t. trace-norm, that is,

\begin{equation}
    \|\Phi(X)\|_1 \leq \|X\|_1 ,
\end{equation}
for all $X^\dagger = X$. Hence, the problem of characterizing semigroups of positive, trace-preserving maps may be equivalently reformulated in terms of contractive semigroups.  Using the theory of Lumer and Philips \cite{Lumer1,Lumer2} of contractive semigroups on Banach spaces he introduces an appropriate semi-inner product $(\!(x,y)\!)$ in $\mathcal{B}_1(\mathcal{H})$ and proved the  following 

\begin{Theorem} A bounded linear operator $\mathcal{L}$ on $\mathcal{B}_1(\mathcal{H})$ generates a dynamical semigroup iff

\begin{enumerate}
    \item ${\rm Re}\, (\!( \rho,\mathcal{L}(\rho) )\!) \leq 0 $ for all $\rho \in \mathcal{B}_1(\mathcal{H})$,
    \item ${\rm Tr}\, \mathcal{L}(\rho)=0$ for all $\rho \in \mathcal{B}_1(\mathcal{H})$.
\end{enumerate}
    
\end{Theorem}
According to Lumer a generator satisfying $(\!( \rho,\mathcal{L}(\rho) )\!) \geq 0 $ is called {\em dissipative}. Hence, any semigroup of positive and trace-preserving maps is generated by a dissipative generator. Finally, Kossakowski provided the following characterization of dissipative generators \cite{AK-1}

\begin{Theorem}\label{TH-72} A bounded linear operator $\mathcal{L}$ on  $\mathfrak{T}(\mathcal{H})$ generates a dynamical semigroup $\Phi_t$ if and only if for every resolution of identity $P_i P_j = \delta_{ij} P_i$, ${\rm tr} P_i < \infty$, and  $\sum_i P_i = \mathbb{I}_\mathcal{H}$ one has
\begin{equation}\label{PLP}
  {\rm tr}( P_i \mathcal{L}(P_j) ) \geq 0  , \ \ \ i \neq j ,
\end{equation}
and $\sum_i  {\rm tr}( P_i \mathcal{L} P_j ) = 0$.
\end{Theorem}
Note, that introducing a real matrix $K_{ij} = {\rm tr}( P_i \mathcal{L} P_j ) $, the above conditions are simply Kolmogorov conditions for a classical Markov generator $K_{ij}$. Observe, that (\ref{PLP}) essentially reduces to the following condition: for any pair of orthogonal vectors $\psi,\phi \in \mathcal{H}$ 

\begin{equation}   \label{cond-positive}
    \< \psi |\mathcal{L}(|\phi\>\<\phi|)|\psi\> \geq 0 . 
\end{equation}
Equivalently, for any rank-1 projector $P$ one has

\begin{equation}  \label{PLP-2}
    P_\perp \mathcal{L}(P) P_\perp \geq 0 ,
\end{equation}
where $P_\perp = \oper_\mathcal{H} - P$ denotes the orthogonal complement of $P$. This property is weaker than positivity which requires (\ref{cond-positive}) for all $\psi,\phi \in \mathcal{H}$. Following \cite{Evans-1,Evans-2} one calls such operators {\em conditionally positive} meaning that  $\mathcal{L}(P)$ is positive but only on a subspace orthogonal to $P$. 
These results
do not assume complete positivity and therefore describe a larger class than
GKLS dynamics.  Their role in the path to the 1976 theorem is nevertheless
fundamental: conditional positivity of $\mathcal{L}$ is the starting point, and replacing
it by conditional \emph{complete} positivity produces the universal GKS
coefficient-matrix structure. 

\begin{Remark} Interestingly, Kossakowski condition (\ref{cond-positive}) is purely geometrical and may be considered as a special case of the following general scenario: consider a convex cone $C$ in $\mathbb{R}^n$ and let $C^\circ$ be a dual cone defined by

\[  C^\circ = \{ y \in \mathbb{R}^n \ | \ (x,y) \geq 0 \ , \ \mbox{for all}\ x \in C\} . \]
Schneider and Vidyasagar\cite{SV} showed that if $B$ is an operator on $\mathbb{R}^n$, i.e. $n \times n$ real matrix, 
then the semigroup  $\{A_t = e^{tB}\}_{t \geq 0}$ satisfies $A_t(C) \subset C$ if and only if 

\[  (y,Bx) \geq 0 ,  \]
for all $x \in C$ and $y \in C^\circ$ such that $(x,y)=0$. Now, $\mathcal{B}_1(\mathcal{H})$ defines a self-dual convex cone in $\mathbb{R}^{N^2}$, where $N = {\rm dim}\, \mathcal{H}$, and $\mathcal{L}$ maybe represented as a real $N^2 \times N^2$ matrix. The result (\ref{cond-positive}) immediately follows. {The} authors of \cite{SV} call such matrices {\rm cross-positive}. 
Hence, conditionally positive operators in $M_N(\mathbb{C})$ correspond to cross-positive matrices in $\mathbb{R}^{N^2}$.

\end{Remark}

In the companion paper on the {\em Quantum Statistical Mechanics of non-Hamiltonian
Systems} \cite{AK-1972}, Kossakowski analyzed one-parameter convolution semigroups of  Markov processes on locally compact groups. As a result he provided  constructions of dynamical semigroups for the harmonic
oscillator.  With
$[a,a^\dagger]=\oper$ and $N=a^\dagger a$, four of his generators are \cite{AK-1972}:

\begin{equation*}
 \mathcal L_1(\rho)=-i[\omega N,\rho]
 -\frac{\sigma}{2}[N,[N,\rho]]
 -\eta[a,[a^\dagger,\rho]] , 
 \tag{110 in Ref.~[14]}
\end{equation*}
\begin{equation*}
 \mathcal L_2(\rho)={}-i[\omega N,\rho]
 -\frac{\sigma}{2}[N,[N,\rho]] 
+\kappa\Bigl([a^\dagger\rho,a]+[a^\dagger,\rho a]\Bigr),
 \tag{116 in Ref.~[14]}
\end{equation*}

\begin{equation*}
 \mathcal L_3(\rho)={}-i[\omega N,\rho]
 -\frac{\sigma}{2}[N,[N,\rho]]\\
 +\delta\Bigl([a\rho,a^\dagger]+[a,\rho a^\dagger]\Bigr),
 \tag{122 in Ref.~[14]}
\end{equation*}
and

\begin{equation*}
 \mathcal L_0(\rho)={}-i[\omega N,\rho]
 -\frac{\sigma}{2}[N,[N,\rho]]-\eta[a,[a^\dagger,\rho]]\\
 +\delta\Bigl([a\rho,a^\dagger]+[a,\rho a^\dagger]\Bigr),
 \tag{128 in Ref.~[14]}
\end{equation*}
where the parameters appearing in the dissipative terms are non-negative.
Formally, all four
have the GKLS structure, although the creation and annihilation operators are
unbounded and the domains must be treated with care.  Complete positivity was
not invoked in his analysis. Kossakowski instead obtained positivity from the explicit
semigroup construction.  He observed that these examples establish positivity
for most master equations then used in laser theory and closed the paper with
the following comment:

\begin{quote}\itshape
``The theory of dynamical semigroups developed so far is still incomplete.
There are many serious mathematical problems to be solved. However, one may
hope that this approach will be useful in a description of open physical
systems.''
\end{quote}

\begin{Remark} In 1973 Kossakowski solved the positivity problem explicitly for a spin
$1/2$ system \cite{AK-2}. This 2-level quantum system was later analyzed in the seminal GKS paper (see section III of \cite{GKS}) where the authors clarified the difference between positivity and complete positivity. From a historical perspective, Kossakowski's 1972-1973 works \cite{AK-1,AK-2} established the geometric and semigroup-theoretic foundations of quantum Markovian dynamics. They introduced conditional positivity, connected quantum generators with Kolmogorov's characterization of classical Markov processes, and provided the conceptual framework later completed by the Gorini-Kossakowski-Sudarshan and Lindblad structure theorems.

\end{Remark}

\subsection{Davies: from microscopic weak coupling to a quantum dynamical semigroup (1974)}
\label{sec:Davies}

The results discussed so far concern the structure and positivity of proposed
master equations.  Davies addressed the complementary problem: when does a
Markovian master equation actually follow from a reversible microscopic
dynamics?  In his 1974 paper \emph{Markovian Master Equations} \cite{Davies1974}, he gave a
rigorous weak-coupling limit for a finite-dimensional system interacting with
an infinite heat bath.  The principal achievement was not another
phenomenological generator, but a proof that reservoir memory disappears on
the correct long time scale and that the reduced dynamics converges to a
quantum dynamical semigroup.

Consider a system $S$ and a reservoir $R$ initially in the product state, and the Hamiltonian

\begin{equation*}
 H_\lambda=H_S+H_R+\lambda H_I,\qquad
 H_I=Q\otimes R , \qquad
 \rho_{SR}(0)=\rho\otimes\sigma_\beta 
 \label{eq:davies-microscopic}
\end{equation*}
where $\sigma_\beta$ is an equilibrium state of the reservoir and
$\operatorname{Tr}_R(R \sigma_\beta)=0$. The exact reduced evolution reads
\begin{equation}
    \label{wcl1}
 \Phi_t^{(\lambda)}(\rho)=
 \operatorname{Tr}_R\!\left[
 e^{-iH_\lambda t}(\rho\otimes\sigma_\beta)e^{iH_\lambda t}\right].
\end{equation}
For every fixed $\lambda\neq0$ this evolution generally displays memory effects and does
not satisfy a semigroup law.  Its first non-trivial dissipative contribution
is of order $\lambda^2$; consequently the limit $\lambda\to0$ at fixed
microscopic time gives only the free unitary system dynamics $\rho_t = e^{-i H_S t} \rho\, e^{i H_S t}$.  Davies therefore used the van Hove \cite{vanHove1955} scaling $\tau=\lambda^2t$ and defined a new dynamical map as a function of $\tau$ (in the interaction picture w.r.t. $H_S$)

\begin{equation}
 \widetilde{\Phi}_{\tau}^{(\lambda)}(\rho) :=
 e^{iH_S\tau/\lambda^2}
 \Phi_{\tau/\lambda^2}^{(\lambda)}(\rho)
 e^{-iH_S\tau/\lambda^2}.
 \label{eq:davies-scaling}
\end{equation}
Thus a finite macroscopic time $\tau$ probes long microscopic
times $t$ as the interaction becomes weaker.

\begin{Theorem}[Davies, 1974]
Let the system be finite-dimensional and suppose that the reservoir
correlation functions satisfy the integrability and decay hypotheses of
Davies's paper \cite{Davies1974}.  Then there exists a bounded generator $\mathcal{K}$ such that,
for every $\tau_1<\infty$,

\begin{equation}
 \lim_{\lambda\to0}\;
 \sup_{0\leq\tau\leq\tau_1}
 \left\|\widetilde{\Phi}_{\tau}^{(\lambda)}
       -e^{\tau\mathcal{K}}\right\|_{1}=0 .
 \label{eq:davies-limit}
\end{equation}
The limiting family $\{e^{\tau\mathcal{K}}\}_{\tau\geq0}$ is a
trace-preserving positive semigroup.  Equivalently, the exact non-Markovian
reduced evolution becomes Markovian on the rescaled time variable.
\end{Theorem}
The structure of $\mathcal{K}$ is obtained by decomposing the coupling
operator $Q$ into its Bohr-frequency components.  If
$H_S=\sum_\varepsilon\varepsilon P_\varepsilon$, define

\begin{equation*}
 e^{iH_St}Qe^{-iH_St}
 =\sum_\omega e^{-i\omega t}A_\omega,\qquad
 A_\omega=
 \sum_{\varepsilon'-\varepsilon=\omega}
 P_\varepsilon QP_{\varepsilon'},\qquad
 A_\omega^\dagger=A_{-\omega}.
 \label{eq:davies-bohr}
\end{equation*}
Let $h(t)=\operatorname{Tr}_R(R(t) R\sigma_\beta)$ be the reservoir
two-point function and
$G(\omega)=\int_{-\infty}^{\infty}e^{i\omega t}h(t)\,dt$.
Bochner’s theorem implies $G(\omega)\geq0$. 
The limiting generator can then
be written in the following form

\begin{equation*}
 \mathcal{K}(\rho)=-i[H_{\mathrm{LS}},\rho]
 +\sum_\omega G(\omega)\left(
 A_\omega\rho A_\omega^\dagger
 -\frac12\{A_\omega^\dagger A_\omega,\rho\}\right),
 \label{eq:davies-generator}
\end{equation*}
and $H_{\rm LS}$ contains the Lamb shift correction to $H_S$, and it  satisfies $[H_S,H_{\mathrm{LS}}]=0$.
This is a GKLS generator resolved into independent Bohr-frequency sectors.
Davies's time averaging is crucial: without it, terms
$A_\omega\rho A_{\omega'}^\dagger$ with $\omega\neq\omega'$ would remain and
the resulting Redfield-type equation \cite{Redfield1957, Redfield1965} need not preserve positivity.  Davies
stated and independently proved that the limit semigroup preserves positivity
and trace.  Complete positivity was not his organizing concept; nevertheless,
each exact reduced map {in Eq.~\eqref{wcl1} as well as in Eq.~\eqref{eq:davies-scaling}} is completely positive,
and the {latter} finite-dimensional norm limit retains this property.  Hence the
limiting family is a quantum dynamical semigroup in the modern sense.

For a thermal reservoir the correlation spectrum satisfies the KMS relation (cf. Section \ref{S:Alicki} for more details) 

\begin{equation}
 G(-\omega)=e^{-\beta\omega}G(\omega)
 \qquad(\omega>0).
 \label{Davies-KMS}
\end{equation}
In this case diagonal density matrices are invariant under
$e^{\tau\mathcal{K}}$ and evolve as a classical continuous-time Markov chain.

\begin{Remark}
    \label{rem:DumckeSpohn}
    The origin of the non-positivity features of generic Redfield-type dynamics and the cure of such a drawback by the 
    Davies time-average prescription are presented in~\cite{DumckeSpohn1979}. As discussed in~\cite{GoriniKossakowski1976}, respectively in~\cite{Dumcke1985}, semigroups of CP maps via coupling to suitable environments and on the correspondingly emerging time-scales (see~\cite{Palmer1977}) can also be obtained  through the singular-coupling, respectively the low-density limits.
\end{Remark}

\begin{Remark} 
The weak-coupling derivation leading to the Davies generator has since become standard material in the theory of open quantum systems and is now presented, at different levels of mathematical rigor, in many monographs and review articles; see, for example, Refs.~\cite{Davies1976,Carmichael1993,Gardiner_Zoller,BreuerPetruccione2002,Attal2006, Rivas-Huelga2011,Schaller2014,Weiss2012, Banerjee2018, Fabio_Roberto, Lidar2019, PR_2022, Szankowski2023,  Vacchini2024}. See also \cite{Tasaki2007, Yuasa2007, Facchi2017} and  a recent pedagogical review in \cite{Stefanini2026}.
\end{Remark}

\begin{Remark}
    In a finite-system/free-boson model, Merkli \cite{Merkli2020} proved, under suitable regularity and Fermi-golden-rule assumptions, an $O(\lambda^2)$ Davies-semigroup error uniformly for all $t\ge 0$, rather than only for bounded $\lambda^2t$. He also obtained an asymptotically exact completely positive semigroup with the correct interacting equilibrium state.
\end{Remark}

\begin{Remark}
A practically important limitation of the standard Davies generator concerns
nearly degenerate Bohr frequencies. The full secular approximation removes all
terms coupling different frequencies $\omega\neq\omega'$. 
This may be too drastic when $|\omega-\omega'|$ is comparable to the 
dissipative rates. Trushechkin \cite{Trushechkin2021} proposed in such a case a  partial secular approximation. 
The resulting unified master equation
retains the relevant nonsecular terms, has GKLS form, and preserves the
appropriate thermodynamic properties.  It reduces to the usual Davies generator when all distinct
Bohr frequencies are well separated. A similar idea was considered by Cattaneo \emph{et al.} who  demonstrated that a global master equation
with a consistent partial secular approximation, retaining the slowly
oscillating cross terms, generally provides a more accurate description than
either the fully secular global equation or a phenomenological local equation
within the Born--Markov regime \cite{Cattaneo2019}.

\end{Remark}

\section{GKLS: the structure theorems (1976) }

\subsection{Gorini-Kossakowski-Sudarshan theorem}

The positivity criterion derived by Kossakowski in \cite{AK-1,AK-2} provides the conceptual starting point of the Gorini-Kossakowski-Sudarshan (GKS) analysis \cite{GKS}. Recall that a generator $\mathcal{L}$ of a positive trace-preserving semigroup satisfies the conditional positivity condition

\[ P_{\perp}\mathcal L(P)P_{\perp}\ge 0 , \] 
for every rank-one projector \(P\), where \(P_{\perp}=\mathbb{I}-P\). Equivalently, for any pair of orthogonal vectors \(|\phi\rangle,|\psi\rangle\in\mathcal{H}\), 
\[ \langle\phi|\mathcal L(|\psi\rangle\langle\psi|)|\phi\rangle\ge 0. \] 
This observation suggests a natural extension to \(k\)-positivity: a generator $\mathcal{L}$ gives rise to a semigroup of \(k\)-positive maps if and only if \(\mathcal{L}\otimes\mathrm{id}_k\) satisfies the corresponding conditional positivity condition. Consequently, \(\mathcal{L}\) generates a completely positive semigroup if and only if it is conditionally completely positive, namely 

\begin{equation} \label{k-cond-positive}
\langle\Phi|(\mathcal L\otimes \mathrm{id}_N) (|\Psi\rangle\langle\Psi|) |\Phi\rangle \ge 0 ,     
\end{equation}
for all mutually orthogonal vectors \(|\Phi\rangle,|\Psi\rangle\in\mathcal{H}\otimes\mathbb{C}^N\).
The breakthrough achieved by Gorini, Kossakowski and Sudarshan was the complete characterization of all generators satisfying this condition. Their result transformed the problem of complete positivity from an infinite family of inequalities into the positivity of a single matrix, now known as the \emph{Kossakowski matrix}. In this way, the authors obtained the first general structure theorem for Markovian quantum dynamics in finite dimensions.

\begin{Theorem} \label{THM-GKS} Let $\{F_\alpha\}_{\alpha =0}^{N^2-1}$ be orthonormal basis in $M_N(\mathbb{C})$ (w.r.t. Hilbert-Schmidt inner product) such that $F_0 = \oper/\sqrt{N}$. Then $\mathcal{L} : M_N(\mathbb{C})\to M_N(\mathbb{C})$ generates a semigroup of completely positive trace-preserving maps if and only if

\begin{equation}   \label{GKS!}
    \mathcal{L}(\rho) = - i[H,\rho] + \frac 12 \sum_{\alpha,\beta=1}^{N^2-1} c_{\alpha\beta} \Big( [F_\alpha,\rho F_\beta^\dagger] + [F_\alpha\rho, F_\beta^\dagger] \Big) , 
\end{equation}
where $H^\dagger = H$, and $c_{\alpha\beta}$ is a complex positive definite matrix (so called Kossakowski matrix). 
\end{Theorem}
Interestingly, based on the Choi theorem \cite{Choi75} who proved that a map $\Phi$ is completely positive if and only if 

\[   ({\rm id}_N \otimes \Phi)(P^+) \geq 0 ,\]
where $P^+$ is {the totally symmetric and} maximally entangled projector in $\mathcal{H} \otimes \mathcal{H}$, one can easily prove that $\mathcal{L}$ is conditionally completely positive if and only if 

\begin{equation}  \label{PLP+}
    P^+_\perp [{\rm id}_N \otimes\mathcal{L}](P^+) P^+_\perp \geq 0 .
\end{equation}
{Namely,  
it is miraculously} sufficient to check (\ref{k-cond-positive}) for a single projector only! Observe, that diagonalizing $c_{\alpha\beta}$ one can transform (\ref{GKS!}) to the well known diagonal form

\begin{equation}   \label{GKS}
    \mathcal{L}(\rho) = - i[H,\rho] +  \sum_{\alpha=1}^{N^2-1} c_{\alpha} \Big( L_\alpha\rho L_\alpha^\dagger - \frac 12 \{L^\dagger_\alpha L_\alpha,\rho\} \Big) , 
\end{equation}
where $c_\alpha \geq 0$ are eigenvalues of $c_{\alpha\beta}$, and $L_\alpha$ are traceless and mutually orthogonal.  It has exactly the same form as the one derived by Bausch \cite{Bausch1966} (cf. eq. (\ref{eq:bausch-generator})) but with an additional constraint for rates $c_\alpha \geq 0$.    

\begin{Remark} A simple and elegant derivation of (\ref{GKS!}) was provided by Alicki \cite{Alicki-Lendi}. Assuming that a semigroup $\Phi_t = e^{t \mathcal{L}}$ consists of completely positive trace-preserving maps any such map can be characterized by its Kraus representation 

\[  \Phi_t(\rho) = \sum_{\alpha,\beta=0}^{N^2-1} a_{\alpha\beta}(t) F_\alpha \rho F_\beta^\dagger , \]
with semipositive definite time dependent matrix $a_{\alpha\beta}(t)$. Initial conditions enforce $a_{00}(0) = N$ and the rest of the matrix elements vanish at $t=0$. Now, the corresponding generator is defined by
$ \mathcal{L} = \frac{d}{dt} \Phi_t|_{t=0}$, 
and simple algebra leads to (\ref{GKS!}) with $c_{\alpha\beta} = \frac{d}{dt}{a}_{\alpha\beta}(t=0)$. Note, that essentially this is the same derivation as the one used in \cite{Bausch1966}. 
\end{Remark}

\begin{Remark}
    In a recent paper, Gen Kimura \cite{Kimura2026} followed the steps of the seminal GKS paper \cite{GKS} to provide a detailed, step-by-step derivation of the GKLS master equation (see also \cite{Legacy, Pascazio2021}).
\end{Remark}



\subsection{Lindblad theorem}

Lindblad's route to the structure theorem [2] was independent of, and conceptually
different from, the finite-dimensional analysis of Gorini, Kossakowski and
Sudarshan \cite{GKS}.  He worked in the Heisenberg picture on a $W^*$-algebra
$\mathcal A$ of observables (essentially $\mathcal{A}=\mathcal B(\mathcal H)$).  His starting axioms were normality, positivity, preservation of the identity, the semigroup law, and ultraweak continuity at
$t=0$.  To obtain an explicit structure theorem he then imposed two additional
assumptions: norm continuity, so that $\Phi_t^\ddag =e^{t\mathcal L^\ddag}$ with a
bounded normal generator, and complete positivity.  Thus, for
$\mathcal A=\mathcal B(\mathcal H)$,

\begin{equation}
  \Phi^\ddag_t\Phi^\ddag_s=\Phi^\ddag_{t+s},\qquad
 \Phi_t^\ddag(\oper)=\oper,\qquad
 \Phi_t^\ddag \otimes\operatorname{id}_n\ \hbox{is positive for every }n.
 \label{eq:lindblad-qds}
\end{equation}
A distinctive feature of Lindblad's analysis is the physical justification of complete positivity. Consider a system \(S_1\) undergoing irreversible evolution and an arbitrary auxiliary system \(S_2\) which does not interact with \(S_1\). The extended evolution must preserve positivity of all states of the compound system \(S_1+S_2\). Since \(S_2\) may have arbitrary finite dimension, Lindblad argued that positivity alone is not sufficient and that one must require complete positivity of the dynamical maps \cite{L}. This argument parallels the motivation given by Gorini, Kossakowski and Sudarshan, although it is developed in a more general operator-algebraic setting.

The key technical tool introduced by Lindblad is the \emph{dissipation form}


\begin{equation}
 \mathfrak{D}_{\mathcal{L}^{\ddag}}(X,Y)
 :=\mathcal{L}^{\ddag}(X^\dagger Y)
  -\mathcal{L}^{\ddag}(X^\dagger)Y
  -X^\dagger\mathcal{L}^{\ddag}(Y) ,
 \label{eq:lindblad-dissipation}
\end{equation}
which quantifies the deviation from reversible Hamiltonian dynamics. Indeed, for a Hamiltonian dynamics --  $\mathcal{L}^\ddag(X)=i[H,X]$ --  one has $\mathfrak{D}_{\mathcal{L}^{\ddag}}(X,Y)=0$, so the dissipation form vanishes exactly for unitary evolution. The relevance of
this form follows by differentiating at $t=0$ the Kadison--Schwarz inequality

\begin{equation}    \label{eq:lindblad-schwarz}
    \Phi^\ddag_t(XX^\dagger) \geq \Phi^\ddag_t(X) \Phi^\ddag_t(X^\dagger) .
\end{equation}

\begin{Theorem}[Lindblad, 1976]
Let $\mathcal A$ be a unital $C^*$-algebra and let
$\mathcal{L}^\ddag: \mathcal A\to\mathcal A$ be a bounded $*$-map satisfying
$\mathcal L^\ddag(\oper)=0$.  Then
$\Phi_t^\ddag=e^{t\mathcal L^\ddag}$ is unital and satisfies
Eq.~\eqref{eq:lindblad-schwarz} for every $t\geq0$ if and only if

\[
 \mathcal{D}_{\mathcal L^\ddag}(X,X)\geq0
 \qquad\text{for every }X\in\mathcal A.
\]
Moreover, $\{\Phi_t^\ddag\}_{t\geq0}$ is a completely positive unital semigroup if
and only if every amplification
$\operatorname{id}_n\otimes\mathcal L^\ddag$ has a positive dissipation form (equivalently, $\mathcal L^\ddag$ is \emph{completely dissipative}).
\end{Theorem}

\begin{Remark} Lindblad called $\mathcal{L}^{\ddag}$ satisfying $\mathfrak{D}_{\mathcal{L}^{\ddag}}(X,X)\geq 0$  {a dissipative generator}. Note, that the notion of {\em dissipativity} used by Lindblad is more restrictive than the one used by Gorini, Kossakowski and Sudarshan. Interestingly, the condition  $\mathfrak{D}_{\mathcal{L}^{\ddag}}(X,X)\geq 0$ restricted to Hermitian operators $X^\dagger=X$ is equivalent to conditional positivity, i.e. $(\oper - P) \mathcal{L}^\ddag(P)(\oper-P) \geq 0$ for all rank-1 projectors $P$.  In finite dimension, complete dissipativity in the Heisenberg picture is the counterpart of
conditional complete positivity of the trace-preserving Schr\"odinger-picture generator.
\end{Remark}
Lindblad proved \cite{L} that every bounded normal completely
dissipative generator can be written as
\[
 \mathcal L^\ddag(X)=  i[H,X] + \Psi(X)-\frac12\{\Psi(\oper),X\},
\]
where  $\Psi$ is {normal and completely positive}.
Conversely, this formula defines a completely dissipative generator on an
arbitrary unital $C^*$-algebra.  {Applying the Kraus-Stinespring-Sudarshan 
representation~\eqref{KSSrep}} gives the familiar operator-sum form.

\begin{Theorem}[Lindblad, 1976]
Let $\mathcal H$ be a separable Hilbert space and let
$\mathcal L^\ddag: \mathcal B(\mathcal H)\to\mathcal B(\mathcal H)$ be bounded
and normal.  It generates a norm-continuous semigroup of normal completely
positive unital maps if and only if there exist
$H=H^\dagger\in\mathcal B(\mathcal H)$ and a finite or countable family
$\{V_j\}\subset\mathcal B(\mathcal H)$, with
$\sum_jV_j^\dagger V_j$ strongly convergent to a bounded operator, such that

\begin{equation}
 \mathcal L^\ddag(X)=i[H,X]
 +\sum_j\left(V_j^\dagger X V_j
 -\frac12\{V_j^\dagger V_j,X\}\right).
 \label{eq:lindblad-heisenberg}
\end{equation}
\end{Theorem}

Indeed, the Hamiltonian part does not contribute to the dissipation form,
whereas the remaining part gives

$$ \mathfrak{D}_{\mathcal{L}^{\ddag}}(X,X)
=\sum_j[X,V_j]^\dagger[X,V_j]\geq0 . $$
Passing to the predual semigroup on trace-class operators,
$\operatorname{Tr}\!\left(\rho\,\mathcal{L}^{\ddag}(X)\right)
=\operatorname{Tr}\!\left(\mathcal{L}(\rho)X\right)$, one obtains the
Schr\"odinger-picture generator

\begin{equation}
 \mathcal{L}(\rho)
 =-i[H,\rho]+\sum_j\left(V_j\rho V_j^\dagger
 -\frac12\{V_j^\dagger V_j,\rho\}\right)
 =-i[H,\rho]+\frac12\sum_j
 \left([V_j\rho,V_j^\dagger]+[V_j,\rho V_j^\dagger]\right).
 \label{eq:lindblad-schrodinger}
\end{equation}

\begin{Remark}
{For $\dim\mathcal{H}=N$, setting
$V_\alpha=\sqrt{c_\alpha}\,L_\alpha$ transforms
\eqref{eq:lindblad-schrodinger} exactly into (\ref{GKS}); conversely, expanding the
$V_j$ in a traceless Hilbert--Schmidt basis recovers the Kossakowski-matrix
form (\ref{GKS!}).  Thus the GKS and Lindblad results are equivalent in finite
dimension, but their formulations are complementary: \cite{GKS} gives a canonical
matrix representation for an $N$-level system, whereas \cite{L} treats
$\mathcal{B}(\mathcal{H})$ in an operator-algebraic framework and allows
infinite-dimensional $\mathcal{H}$ under the bounded-generator assumption (unbounded generators were studied e.g. in Ref. \cite{Holevo-Werner}).
Both Lindblad and GKS explicitly cited Davies’s weak coupling results \cite{Davies1974}. {Lindblad's} manuscript was received on
April 7, 1975, revised on November 28, 1975, and published in 1976; in a note
added after completion he acknowledged the related preprint by Gorini,
Kossakowski and Sudarshan.}

\end{Remark}

\begin{Remark}
    A further interesting contribution appeared in 1976 in a paper by
Franke \cite{Franke1976}.
Apparently unaware of the works of Gorini, Kossakowski, and
Sudarshan and of Lindblad, Franke studied trace-preserving semigroups
consisting of positive decomposable maps. Such maps can be represented
as sums of a completely positive and a completely copositive map, and
the corresponding generator contains, besides the usual GKLS part, an
additional contribution involving transposition. For a two-dimensional
systems (qubits), where every positive map is decomposable, Franke obtained
the general structure of generators of positive trace-preserving
semigroups; this is closely related to the low-dimensional
decomposability result established by Woronowicz
\cite{Woronowicz}.

The status of Franke's result is nevertheless different from that of
the GKS and Lindblad theorems. The decomposition into completely
positive and completely copositive parts is highly nonunique, and
positivity or decomposability does not guarantee consistency after
extension by an arbitrary ancilla. It is complete positivity that
provides this stability and therefore leads to the physically standard
GKLS structure. Franke's independent analysis nonetheless deserves
recognition and has motivated the occasional use of the acronym
FGKLS \cite{
FGKLS}.

{
In a profound sense, it is Nature and Physics that have chosen to follow the path of complete positivity: the requirement that the evolution of an open system remain physically consistent in the presence of an arbitrary, non-interacting ancilla is not a mathematical convenience but an inescapable physical demand, and it is this demand --- not mere positivity or decomposability --- that singles out the GKLS structure as the correct and universal framework.}

\end{Remark}

\section{After GKLS: early developments}

\subsection{Alicki detailed balance generators (1976)}   \label{S:Alicki}

In his 1976 paper on non-Hamiltonian systems, Alicki \cite{Alicki_1976} analyzed quantum
detailed balance for finite-dimensional dynamical semigroups (it was a subject of his master thesis supervised by Andrzej Kossakowski). The paper \cite{Alicki_1976} was received on
March 3, 1976 and already cited the then forthcoming GKS theorem as the general
form of a completely positive generator.   The definition adopted in \cite{Alicki_1976}, proposed by Kossakowski, is most
naturally formulated in the Heisenberg picture.  Let $\rho_0>0$ be a faithful
state and equip $\mathcal{B}(\mathcal{H})$ with the weighted Hilbert--Schmidt
inner product

\begin{equation}
 \langle A,B\rangle_{\rho_0}
 :=\operatorname{Tr}(\rho_0A^\dagger B).
 \label{eq:alicki-inner-product}
\end{equation}
Denote by $\mathcal{L}^{\ddag}$ the GKLS generator in the Heisenberg picture and by $\widetilde{\mathcal{L}^{\ddag}}$ the adjoint of the Heisenberg
generator $\mathcal{L}^{\ddag}$ with respect to this inner product, that is,

\[   \langle \mathcal{L}^{\ddag}(A),B\rangle_{\rho_0} =  \langle A,\widetilde{\mathcal{L}^{\ddag}}(B)\rangle_{\rho_0} . \]
The quantum detailed-balance condition with respect to $\rho_0$ is defined as follows \cite{Alicki_1976}:

\begin{enumerate}
    \item there exists $H=H^\dagger$ such that the anti-Hermitian part of $\mathcal{L}^{\ddag}$ is a Hamiltonian derivation

\begin{equation}
  \frac 12 \Big( \mathcal{L}^{\ddag}(X) - \widetilde{\mathcal{L}^{\ddag}}(X) \Big)  =  i[H,X] ,   
\end{equation}

\item ${\mathcal{L}^{\ddag}}$ is normal w.r.t. (\ref{eq:alicki-inner-product}), that is, $ [\mathcal{L}^{\ddag},\widetilde{\mathcal{L}^{\ddag}}]=0$.
 
\end{enumerate}
 This formulation contains both classical
reversibility, where the Markov generator is self-adjoint in the stationary
weighted inner product, and Hamiltonian evolutions, whose generator is
anti-self-adjoint.  Now, if the generator $\mathcal{L}^\ddag$ satisfies quantum detailed balance condition w.r.t. $\rho_0$, then $\rho_0$ is stationary for the predual (Schr\"odinger picture)
semigroup, i.e. $\mathcal{L}(\rho_0) = 0$. Moreover, when $\rho_0$ is non-degenerate, diagonal and  off-diagonal matrix elements of $\rho(t)$ in the  eigenbasis of $\rho_0$ evolve independently.

Consider the spectral representation of $\rho_0$, i.e. $\rho_0=\sum_{k=1}^Nr_k |k\>\<k|$, with $r_k>0$, and assume that the 
anti-Hermitian part of the generator is $i[h,\,\cdot\,]$, where
$[h,\rho_0]=0$, while the Hamiltonian and dissipative parts commute. Finally, let us define

\begin{equation}
    \Delta(X) = \rho_0 X \rho_0^{-1} .
\end{equation}
    
\begin{Theorem}[Alicki, 1976]
\textit{Under the preceding assumptions there exist operators $X_{ij}$ and
numbers $D_{ij}\geq0$ such that}

\begin{equation}
 \mathcal{L}^{\ddag}(A)=i[h,A]
 +\sum_{i,j=1}^{N}D_{ij}
 \left(X_{ij}^\dagger[A,X_{ij}]+[X_{ij}^\dagger,A]X_{ij}\right),
 \label{eq:alicki-paired-generator}
\end{equation}
\textit{where the jump operators $X_{ij}$ and rates $D_{ij}$ may be chosen to satisfy}

\begin{equation}
 \Delta(X_{ij})=\frac{r_i}{r_j}X_{ij},\qquad
 X_{ij}^\dagger=X_{ji},\qquad
 \operatorname{Tr}(X_{ij}^\dagger X_{kl})=\delta_{ik}\delta_{jl},
 \qquad
 D_{ij}r_j=D_{ji}r_i.
 \label{eq:alicki-pairing}
\end{equation}
    
\end{Theorem}

The last identity is the quantum counterpart of the classical balance of each
forward and backward transition.  Indeed, if
$\rho(t)=\sum_np_n(t) |n\>\<n|$ commutes with $\rho_0$, then the diagonal sector
closes and

\begin{equation}
 \dot p_n(t)=\sum_k\big(W_{nk}\,p_k(t)-W_{kn}\,p_n(t)\big),
 \label{eq:alicki-classical-balance}
\end{equation}
where the rates $   W_{nk} = 2 \sum_{i,j=1}^N D_{ij} |\<n|X_{ij}|k\>|^2 $, 
satisfy $W_{nk}r_k=W_{kn}r_n$. Thus the quantum condition reduces exactly to classical detailed balance
on populations, while the coherences form invariant complementary sectors.

The result takes a particularly transparent form for the Gibbs state
$\rho_\beta=Z^{-1}e^{-\beta H_S}$.  One finds \cite{Alicki_1976}

\begin{equation}
    \mathcal{L}(\rho) = -i[H_{\rm eff},\rho] +  \sum_{\omega \geq 0} \left\{ \gamma_+(\omega)\Big( V_\omega \rho V_\omega^\dagger - \frac 12 \{V_\omega^\dagger V_\omega,\rho\} \Big) + \gamma_-(\omega) \Big(  V_\omega^\dagger \rho V_\omega - \frac 12 \{V_\omega V_\omega^\dagger ,\rho\} \Big) \right\} ,
\end{equation}
where $e^{i H_S t} V_\omega e^{-iH_S t} = e^{-i \omega t} V_\omega$  and $[H_S,H_{\rm eff}]=0$. Finally, the exponential relation between opposite Bohr frequencies 

\begin{equation}
    \gamma_-(\omega) = e^{- \beta \omega}  \gamma_+(\omega) \,
\end{equation}
is the familiar thermal, or KMS, balance relation.  Alicki noted that the same generator had
been obtained by Davies for an $N$-level system weakly coupled to an infinite
heat bath \cite{Davies1974} (as in formula (\ref{Davies-KMS})).

\begin{Remark} 
Detailed balance condition was further analyzed in  \cite{Kossakowski1977, Gorini1978} (compare with the analysis based on the Agarwal's definition \cite{CarmichaelWalls1976}). Extensive analysis of detailed balance w.r.t. to one-parameter family of inner products 

\[   (X, Y )_s :=  \Tr(\rho_0^s X^\dagger \rho_0^{1-s} Y ) ; \quad s\in [0,1] ,\]
was performed e.g. in \cite{FagnolaUmanita2007, FagnolaUmanita2008, CarlenMaas2017}. 
\end{Remark}

\subsection{Spohn-Frigerio theory of relaxation and approach to equilibrium (1976--1977)}


In his 1976 paper \cite{Spohn_1976} Spohn provided a simple sufficient condition for a completely positive dynamical semigroup of a $N$-level system to have a unique (invariant) equilibrium. 
Spohn called the semigroup \emph{relaxing} if there exists a state $\rho_0$
such that

\begin{equation}
 \lim_{t\to\infty}\Phi_t(\rho)=\rho_0
 \qquad\text{for every state }\rho.
 \label{eq:spohn-relaxing}
\end{equation}
In finite dimension this implies that $\rho_0$ is the unique stationary state
and that all non-zero eigenvalues of $\mathcal{L}$ have strictly negative real
parts, so the convergence is exponential. Let $\mathcal{H}$ be $N$-dimensional and let
$\Phi_t=e^{t\mathcal{L}}$ be a completely positive trace-preserving semigroup.
In a traceless Hilbert--Schmidt orthonormal basis
$\{F_\alpha\}_{\alpha=1}^{N^2-1}$ its generator has the canonical GKLS form

\begin{equation}
 \mathcal{L}(\rho)=-i[H,\rho]
 +\sum_{\alpha,\beta=1}^{N^2-1}c_{\alpha\beta}
 \left(F_\alpha\rho F_\beta^\dagger
 -\frac12\{F_\beta^\dagger F_\alpha,\rho\}\right),
 \qquad c=[c_{\alpha\beta}]\geq 0.
 \label{eq:spohn-generator}
\end{equation}

\begin{Theorem}[Spohn, 1976]
If $p$ is the multiplicity of the eigenvalue zero of the 
Kossakowski matrix $c=[c_{\alpha\beta}]$ and

\begin{equation}
 p<\frac{N}{2},
 \label{eq:spohn-nullity}
\end{equation}
then the semigroup $\{\Phi_t\}_{t\geq0}$ is relaxing.  In particular, every strictly
positive Kossakowski matrix generates a relaxing semigroup.
\end{Theorem}
The above condition has a direct reservoir interpretation: sufficiently many
independent system degrees of freedom must be effectively coupled to the bath.
It is only sufficient, however, and can be much stronger than necessary.  In
his 1977 paper \cite{Spohn_1977} Spohn formulated the more intrinsic condition in terms of a
diagonal representation

\begin{equation}
 \mathcal{L}(\rho)=-i[H,\rho]
 +\sum_{j\in J}\left(V_j\rho V_j^\dagger
 -\frac12\{V_j^\dagger V_j,\rho\}\right).
 \label{eq:spohn-lindblad-form}
\end{equation}
Spohn \cite{Spohn_1977} identified irreducibility of the noise algebra as the essential
mechanism behind relaxation.

\begin{Theorem}[Spohn, 1977]    If
$\operatorname{span}_\mathbb{C}\{V_j:j\in J\}$ is self-adjoint and the noise operators act
irreducibly on the Hilbert space $\mathcal{H}$, equivalently {if the double commutant} $ \{V_j:j\in J\}''=\mathcal{B}(\mathcal{H})$, 
%
then the semigroup $\{\Phi_t\}_{t\geq0}$ is relaxing.
\end{Theorem}
The above theorem provides the first simple algebraic criterion guaranteeing relaxation of finite-dimensional GKLS semigroups. Almost simultaneously, Frigerio \cite{Frigerio_1977,Frigerio_1978} reformulated the problem in the language of operator algebras, proving that irreducibility, uniqueness of the invariant state and triviality of the fixed-point algebra are equivalent properties for semigroups possessing a faithful stationary state. He further extended Spohn's convergence result from finite-dimensional systems to a much broader class of quantum dynamical semigroups. Together, these works established the mathematical foundations of quantum relaxation and thermalization for Markovian dynamics.

\begin{Theorem}[Frigerio, 1977]
    Let $\{\Phi_t\}_{t\ge0}$ be a quantum dynamical semigroup possessing a faithful normal stationary state $\rho_0$. Then the following properties are equivalent: 
    
    \begin{enumerate} 
    
    \item $\rho_0$ is the unique stationary state; 
    
    \item the fixed-point algebra is trivial, 
    
    \[ F(\Phi):=\{A\;|\;\Phi^\ddag_t(A)=A,\ \forall\, t\ge0\} = \mathbb{C}\,\mathbf{1}; \] 
    
    \item the semigroup is irreducible, i.e. $\{H, V_j, V_j^\dagger:j \in J\}' = \mathbb{C}\oper$.
    
    \end{enumerate}
\end{Theorem}
The results of Spohn and Frigerio completed an important step in the early development of open quantum systems. The GKLS theorem identifies the allowed generators of completely positive Markovian dynamics, while the Spohn-Frigerio analysis explains when such dynamics actually thermalizes and completely forgets its initial condition. These works established irreducibility and relaxation to equilibrium as central concepts in the modern theory of quantum dynamical semigroups.

\begin{Remark}

Recently, Yoshida \cite{Yoshida_2024} revisited the problem of the uniqueness of nonequilibrium steady states (NESS) for finite-dimensional GKLS generators and provided a particularly simple algebraic proof of a sufficient condition for uniqueness. He showed that if the set \[ \left\{ H-\frac{i}{2}\sum_m L_m^\dagger L_m,\, L_1,\ldots,L_M \right\} \] generates the full operator algebra $B(\mathcal H)$ under addition, scalar multiplication and multiplication, then the stationary state is unique and full-rank. Unlike Frigerio's theorem, which assumes the existence of a faithful stationary state and then characterizes its uniqueness through irreducibility, Yoshida's criterion directly implies both uniqueness and faithfulness of the steady state. His work provides a finite-dimensional and elementary proof of ideas that can be traced back to the seminal results of Spohn \cite{Spohn_1976,Spohn_1977}, Frigerio \cite{Frigerio_1977,Frigerio_1978} and Evans 
\cite{Evans_1977}. See also the recent review \cite{ZhangBarthel2024}. 
For the time-dependent generalization of Spohn-Frigerio result cf. \cite{JMP_2026}.

\end{Remark}

\subsection{Spohn: entropy production and irreversible thermodynamics (1978)}
\label{sec:entropy-production}

The study of stationary states and relaxation was soon complemented by
an entropy-based description of irreversibility. In 1978, Spohn
\cite{Spohn1978} introduced an entropy-production functional for
quantum dynamical semigroups with a stationary state and established its
nonnegativity and convexity. Together with the work of Spohn and Lebowitz
on systems weakly coupled to thermal reservoirs
\cite{SpohnLebowitz1978}, this provided a connection between the structure
of Markovian quantum dynamics and the entropy balance of irreversible
thermodynamics.

Consider a finite-dimensional completely positive, trace-preserving
semigroup $\Phi_t=e^{t\mathcal L}$ with a faithful stationary state
$\rho_0>0$, so that $\mathcal L(\rho_0)=0$. The quantum relative entropy
with respect to $\rho_0$ is defined by
\begin{equation}
 S(\rho\Vert\rho_0)
 =\operatorname{Tr}\bigl[\rho(\log\rho-\log\rho_0)\bigr].
 \label{eq:ep-relative-entropy}
\end{equation}
Then the entropy production (with respect to $\rho_0$) is defined as follows \cite{Spohn1978}

\begin{equation}
 \sigma_{\rho_0}(\rho)
 :=-\left.\frac{d}{dt}S(\Phi_t(\rho)\Vert\rho_0)\right|_{t=0}
 =-\operatorname{Tr}\bigl[\mathcal L(\rho)
             (\log\rho-\log\rho_0)\bigr].
 \label{eq:ep-spohn-functional}
\end{equation}
The explicit logarithmic formula is stated for faithful states; at the
boundary of the state space one can instead use the corresponding
one-sided relative-entropy derivative, which need not be finite.

\begin{Theorem}[Spohn, 1978]
For a finite-dimensional GKLS generator $\mathcal L$ possessing a
faithful stationary state $\rho_0$, one has
\begin{equation}
 \sigma_{\rho_0}(\rho)\geq0
 \label{eq:ep-spohn-inequality}
\end{equation}
for every faithful density operator $\rho$. 
\end{Theorem}

A short proof follows from the monotonicity of quantum relative entropy
under completely positive, trace-preserving maps \cite{Lindblad1975}. Since
$\Phi_t(\rho_0)=\rho_0$, one finds
\begin{equation}
 S(\Phi_t(\rho)\Vert\rho_0)
 =S(\Phi_t(\rho)\Vert\Phi_t(\rho_0))
 \leq S(\rho\Vert\rho_0).
 \label{eq:ep-contraction}
\end{equation}
Differentiating at $t=0$ proves
\eqref{eq:ep-spohn-inequality}; trace preservation eliminates the term
$\operatorname{Tr}\mathcal L(\rho)$ in the derivative. Along a trajectory
$\rho_t=\Phi_t(\rho)$, the same argument gives
\begin{equation}
 \frac{d}{dt}S(\rho_t\Vert\rho_0)
 =-\sigma_{\rho_0}(\rho_t)\leq0.
 \label{eq:ep-relative-decay}
\end{equation}

The thermodynamic meaning becomes explicit when $\rho_0$ is a thermal state, i.e.
$\rho_0=\rho_\beta=Z^{-1}e^{-\beta H_S}$, with $H_S$ time independent system's Hamiltonian.
Set $k_B=1$ and write $S(\rho)=-\operatorname{Tr}(\rho\log\rho)$. Defining the heat current
\begin{equation}
 \dot Q(t)=\operatorname{Tr}\bigl[H_S\mathcal L(\rho_t)\bigr] ,
 \label{eq:ep-heat-current}
\end{equation}
and using $\log\rho_\beta=-\beta H_S-\log Z$ in
\eqref{eq:ep-spohn-functional}, one obtains the entropy balance
\begin{equation}
 \sigma_{\rho_\beta}(\rho_t)
 =\frac{d}{dt}S(\rho_t)-\beta\dot Q(t)\geq0.
 \label{eq:ep-clausius}
\end{equation}
The system entropy  $S(\rho_t)$ can therefore decrease during relaxation, provided
the entropy transferred to the reservoir compensates for that decrease.
When the reference state is maximally mixed, the semigroup is unital and
\eqref{eq:ep-clausius} reduces to $dS(\rho_t)/dt\geq0$.


Spohn and Lebowitz \cite{SpohnLebowitz1978} generalized the above analysis for multi-reservoirs scenario. 
 Suppose
\begin{equation}
 \mathcal L(\rho)=-i[H_{\mathrm{eff}},\rho]
             +\sum_a\mathcal L_a(\rho),
\end{equation}
where every $\mathcal L_a$ is a GKLS generator with the faithful thermal
stationary state $\rho_{\beta_a}=Z_a^{-1}e^{-\beta_a H_S}$, and assume $ [H_{\mathrm{eff}},H_S]=0$. 
Applying Spohn's inequality (\ref{eq:ep-spohn-inequality}) separately to each contribution gives
\begin{align}
 \sigma_{\mathrm{tot}}(\rho_t)
 :=-\sum_a\operatorname{Tr}\bigl[\mathcal L_a(\rho_t)
                  (\log\rho_t-\log\rho_{\beta_a})\bigr]
 =\frac{d}{dt}S(\rho_t)-\sum_a\beta_a\dot Q_a(t)\geq0,
 \label{eq:ep-multiple-reservoirs}
\end{align}
with $\dot{Q}_a(t)=\operatorname{Tr}\bigl[H_S\mathcal L_a(\rho_t)\bigr]$.
At a faithful nonequilibrium stationary state $\rho_{\mathrm{ss}}$,
the heat currents need not vanish and
$\sigma_{\mathrm{tot}}(\rho_{\mathrm{ss}})
=-\sum_a\beta_a\dot Q_a\geq0$ can be strictly positive. By contrast,
the functional constructed from the total generator with
$\rho_{\mathrm{ss}}$ as its reference state vanishes at stationarity.

These results complemented the algebraic criteria for relaxation with
a thermodynamic description of the same dynamics. The GKLS structure
ensures a consistent evolution of states, while the stationary state and,
when relevant, the decomposition into reservoir contributions determine
the associated entropy balance. For recent reviews on irreversible entropy production and other aspects of quantum theormodynamics see e.g. \cite{Kosloff2013, Marcantoni2017, Kosloff2019, LandiPaternostro2021} and monographs \cite{GemmerMichelMahler2009,  DeffnerCampbell2019, Binder2018}.

\subsection{Lindblad: quantum regression and Markovianity (1979)}
\label{sec:QR}

Davies derivation of a quantum dynamical senioroup in Section~\ref{sec:Davies} starts with the so-called reduced dynamics.
Namely, with a family of maps on the states of an open system $S$ in interaction with its environment $R$ 
(see Eq.~\eqref{wcl1}), formally obtained as 
\begin{equation}
    \label{red-dyn}
    \Lambda_t(\rho_S):={\rm Tr}_R\Big(\mathcal{U}_t(\rho_S\otimes\rho_R)\Big)\ .
\end{equation}
The reduced dynamics thus emerges from the unitary dynamics
\begin{equation}
    \label{un-dyn}
    \mathcal{U}_t(\rho_S\otimes\rho_R)=U_{t}\,\rho_S\otimes \rho_R\,U^\dag_t\ ,\quad U_t=\exp(-i\,t\, H_T)\ ,
\end{equation}
generated by the global Hamiltonian $H_T$ of the compound system $S+R$ by eliminating the degrees of freedom
of the environment via the partial trace $\Tr_R$. Certainly, apart from trivial settings, the maps $\Lambda_t$ do not 
obey a semigroup composition law 
\begin{equation}
\label{sem-law}
\Lambda_t\circ\Lambda_s=\Lambda_s\circ\Lambda_t=\Lambda_{t+s}\qquad\forall s,t\geq 0\ .
\end{equation}
However, as discussed in Section~\ref{sec:Davies} through suitable limits and on specific long time-scales, the reduced dynamics $\Lambda_t$ becomes of the exponential form $\exp(t\,\mathcal{L})$ with GKSL generator $\mathcal{L}$. 

Usually, the exponential form $\Lambda_t=\exp(t\mathcal{L})$ and the ensuing time-homogeneity is taken as a signature of Markovianity, namely of memorylessness of the dynamics (see Eq.\eqref{time-hom} in Section~\ref{S:Bausch}).
Instead, as pointed out by Lindblad in~\cite{Lindblad1979}, a semigroup law is
not sufficient for the system dynamics to be Markovian. 
This is true also in a classical setting where the Markov character of a stochastic process, namely the fulfillment of the 
Chapman-Kolmogorov equations, needs the inspection of all its higher order time-correlation functions.

Using the Heisenberg unitary dynamics  $\mathcal{U}_t^\ddagger$ dual to the one in~\eqref{un-dyn}, consider the two-point time-correlations of two observables $A_{1,2}$ of the system $S$, only. Using the cyclicity of the trace, from the group composition law of the unitary maps $\mathcal{U}_t$, these correlation functions can be recast as
\begin{eqnarray*}
\label{qr1}
\langle A_2(t_2)A(t_1)\rangle_S&:=&\Tr_{S+R}\Big(\rho_S\otimes\rho_R\,\mathcal{U}^\ddag_{t_2}\Big(A_2\otimes\mathds{1}_R\Big)\,
\mathcal{U}^\ddag_{t_1}\Big(A_1\otimes\mathds{1}_R\Big)\Big)\\
&=&\Tr_S\Big(\Tr_R\Big(A_2\,\mathcal{U}_{t_2-t_1}\Big(A_1\otimes\mathds{1}_R\Big)\,\mathcal{U}_{t_1}\Big(\rho_S\otimes\rho_R\Big)\Big)\Big)\ .
\end{eqnarray*}
Notice that, setting $A_1=\mathds{1}_S$ and varying $A_2$ arbitrarily, one retrieves the action of the reduced dynamics $\Lambda_t$ in~\eqref{red-dyn}. 

Quantum regression is said to hold true whenever one is allowed to replace with CP maps $\Lambda_{t,s}$ and for each given system observable the unitary propagators and the trace over the environment so that
\begin{equation}
\label{qr2}
\langle A_2(t_2)A(t_1)\rangle_S=\Tr_S\Big(A_2\,\Lambda_{t_2,t_1}\Big(A_1\,\Lambda_{t_1}(\rho_S)\Big)\Big)\ .
\end{equation}
Extrapolating to $n$-point time-correlation functions, quantum regression reads
\begin{equation}
\langle A_n(t_n)\cdots A_2(t_2)A(t_1)\rangle_S=\Tr_S\Big(A_n\,\Lambda_{t_n,t_{n-1}}\Big(A_{n-1}\cdots\Lambda_{t_3,t_2}\Big(A_2\Lambda_{t_2,t_1}\Big(A_1\,\Lambda_{t_1}(\rho_S)\Big)\Big)\Big)\Big)\ .
\end{equation}
By varying freely $A_{2,3}$ while fixing $A_1=A_4,\ldots,A_n=\mathds{1}_S$, one gets the two parameter semigroup 
composition laws
\begin{equation}
\label{non-mark}
\Lambda_{t,0}=\Lambda_t\,,\quad \Lambda_{t,s}\circ\Lambda_{s,u}=\Lambda_{t,u}\qquad\forall t\geq s\geq u\geq 0\ .
\end{equation}
In a commutative setting whereby both observables and states are diagonal with respect to a same orthonormal basis, the
regression structure reproduces the Chapman-Kolomogorov equations for the transition matrices of a classical Markov process. 

This review of the early stages of open quantum dynamics is centered around the GKSL theorem and its early implications and cannot accommodate a topical matter as that concerning  the still developing field of quantum open dynamics with  generators that are explicitly time-dependent, for which we refer to the reviews~\cite{PR_2022, RHP, Vega, BLPV,LiHallWiseman2018}.

We just observe that the CP character of the propagators make the family of maps $\Lambda_t$ CP-divisible and thus Markovian in one of the formal qualifications of this notion present in the literature~\cite{GuarnieriSmirneVacchini2014}.
However, the fact that quantum regression implies but is not implied by CP-divisibility~\cite{Budini2008,GuarnieriSmirneVacchini2014}
confirms the initial observation by Lindblad. Namely, that the one-time reduced dynamics itself is not able to capture the depth of non-Markovianity features as, for instance, the information flows between systems $S$ and environment $R$ characterizing an open memoryfull quantum dynamics. For such a task, multi-time correlation functions must be investigated, very much in the spirit of quantum process tensors or quantum combs proposed recently by Modi and collaborators ~\cite{Pollock2018, MilzModi2021}. Such an approach puts together unitary dynamics and intervening generalized quantum measurement process {as outlined} 
in the seminal paper~\cite{AFL1982}

\subsection{Hudson-Parthasarathy: quantum stochastic calculus (1984)}
\label{sec:HP}

While the reduced dynamics obtained from Eq.~\eqref{red-dyn}, does not in general   provide a semigroup, soon after the formulation  of the GKSL theory,
Evans and Lewis in~\cite{EvansLewis1976} showed that each quantum dynamical semigroup $\Lambda_t$ with bounded can be dilated to a unitary group $U_t$ over a larger compound system $S+R$. Namely, one can construct an environment  Hilbert space $\mathcal{H}_R$ a group of unitaries $U_t$ acting on $\mathcal{H}_S\otimes\mathcal{H}_R$ and an environment state $\rho_R$ 
such that Eq.~\eqref{red-dyn} holds.

In\cite{HudsonParthasarathy1984}, Hudson and Parthasarathy considered a concrete Bosonic environment described by a Fock space $\mathcal{F}$ with creation and annihilation operators of frequency modes
$a(\omega)$, $a^\dag(\omega)$ satisfying the canonical commutation relations
$[a(\omega_1)\,,\,a^\dag(\omega_2)]=\delta(\omega_1-\omega_2)$.  Consider the canonically conjugated  creation and annihilation operators
$$
b(t)=\frac{1}{\sqrt{2\pi}}\int_{\mathbb{R}}{\rm d}\omega\,{\rm e}^{-i\,
\omega\,t}\,a(\omega)\ ,\qquad b^\dag(t)=\frac{1}{\sqrt{2\pi}}\int_{\mathbb{R}}{\rm d}\omega\,{\rm e}^{i\,
\omega\,t}\,a^\dag(\omega)
$$
that satisfy the commutation relations $[b(t_1)\,,\,b^\dag(t_2)]=\delta(t_1-t_2)$.
The Fock space is characterized by a vacuum state $\ket{\Omega}$ such that {$a(\omega)\ket{\Omega}=0$ for all $\omega$ and thus} $b(t)\,\ket{\Omega}=0$ for all $t\in\mathbb{R}$. 
{State vectors in $\mathcal{F}$ are then constructed} by acting with polynomials of every order in the single particle creation operators 
$b^\dag(f)$ such that $b^\dag(f)\ket{\Omega}=\ket{f}$, where $f(t)$ any square-summable over $t\geq 0$.
Then, define so-called noise-operators, namely  annihilators and creators of characteristic functions of time intervals $\chi_{[0,t]}(\tau)$: 
\begin{equation}
    \label{noise-op}
    A_t:=A([0,t])=\int_0^t{\rm d}s\,b(s)\ ,\qquad A^\dag_t:=A^\dag([0,t])=\int_0^t{\rm d}s\,b^\dag(s)\ .
\end{equation}
With $\Delta A_t:=A_{t+\Delta t}-A_t$, one finds
$$
\Delta A_t\,\Delta A^\dag_t=\int_t^{t+\Delta t}{\rm d}s_1\int_t^{t+\Delta t}{\rm d}s_2\,b(s_1)b^\dag(s_2)\, 
=\Delta A^\dag_t\,\Delta A_t+\Delta t\ .
$$
so that, letting $\Delta t\to dt$, the action on the vacuum of an operator product $\Delta A_t\,\Delta A^\dag_t$ satisfies
\begin{equation}
\label{quantumIto}
{\rm d}A_t\,{\rm d}A^\dag_t\ket{\Omega}=dt\ket{\Omega}\ ,
\end{equation}
which is the non-commutative analogue of the relation $(dw(t))^2=dt$ satisfied by the Wiener process $w(t)$.
Such relation is one from a table of Quantum Ito products ~\cite{HudsonParthasarathy1984}
\[ \mathrm dA_t\,\mathrm dA_t^\dagger=\mathrm dt,
  \qquad
  \mathrm dA_t^\dagger\,\mathrm dA_t
  =(\mathrm dA_t)^2
  =(\mathrm dA_t^\dagger)^2=0, 
\]
indicating that, in perturbative expansions with respect to ${\rm d}A_t$ and ${\rm d} A^\dag_t$ one should be careful to keep specific second order terms as for instance in the differential operator stochastic equation
\begin{equation}
\label{adaptedness0}
{\rm d}(X_t\,Y_t)=({\rm d} X_t)\,Y_t\,+\, X_t\,({\rm d} Y_t)
+({\rm d} X_t)({\rm d} Y_t)\ ,
\end{equation}
which holds for adapted quantum stochastic processes $X_t$, $Y_t$.

\begin{Remark}
\label{rem:adapt}
The notion of adaptedness is of central importance and corresponds to an assumption of non-anticipation.
Namely, an operator-valued process $X_t$ is called adapted if it acts as
\begin{equation}
\label{adaptedness1}
X_t=X_t^{[0,t]}\otimes {\rm I}_{[t,\infty)} 
\end{equation}
on the tensor splitting $\mathcal{F}=\mathcal{F}_{[0,t]}\otimes\mathcal{F}_{[t,+\infty)}$ of the Fock space,
where ${\rm I}_{(t,\infty)}$ represents the identity action on all 
states in the component $\mathcal{F}_{[t,+\infty)}$ created by acting on the vacuum with creators of functions supported by any interval of 
time after $t$. 

An adapted process cannot therefore act on future and represent the quantum analogue of the classical idea that a stochastic
process at time $t$ only depends on information available by time $t$.
\end{Remark}

For simplicity, fix a GKSL generator with only one noise operator $L$ and  consider the equation for an infinitesimal operator increment
\begin{equation}
\label{HPdiff}
{\rm d} U_t = \left(L\,{\rm d} A^\dagger\, -\,L^\dagger\,{\rm d} A\,+\,
K\,{\rm d} t\right)\,U_t\ ,\quad K:=\,-\,i\,H\,-\frac{1}{2}L^\dagger L\ ,
\end{equation}
where $U_t$ is an adapted quantum stochastic process.
Then, using~\eqref{adaptedness0} and~\eqref{quantumIto}, one finds
$$
{\rm d}(U^\dag_t U_t)=({\rm d}U^\dag_t)\, U_t\,+\,U^\dag_t\,({\rm d}U_t)\,+\,{\rm d}U^\dag_t\,{\rm d}U_t\ ,
$$
where, using that the system operators and those of the environment commute,
$$
({\rm d}U^\dag_t)\, U_t\,+\,U^\dag_t\,({\rm d}U_t)=U_t^\dag(K+K^\dag)\,U_t\, \text{d}t =-U_t^\dag L^\dag L\,U_t\, \text{d}t 
=-{\rm d}U^\dag_t\,{\rm d}U_t ,
$$
so that ${\rm d}(U^\dag_t\,U_t)=0$ (analogously for ${\rm d}(U_t\,U^\dag_t)=0$) and $U_t$ is a unitary process.

For a system observable $X$, define the embedding at time $t$ {into the system-environment algebra}
$$
j_t(X)=U_t^\dagger(X\otimes I)U_t\ ,
$$
where $U_t$ solves~\eqref{HPdiff}. Then, again by adaptedness, it follows that the embedding satisfy the quantum stochastic differential 
equation
$$
{\rm d}j_t(X)=
({\rm d} U_t^\dagger)\,X\,U_t\,+\,
U_t^\dagger\, X\,({\rm d} U_t)\,+\,
({\rm d} U_t^\dagger)\,X\,({\rm d} U_t)\ .
$$
Then, the quantum Ito calculus yields the quantum Langevin equation
\begin{equation}
\label{quantumLangevin}
{\rm d} j_t(X)=
j_t(\mathcal{L}(X))\,{\rm d}t\,+\,j_t([X,L])\,{\rm d} A_t^\dagger\,+\,
j_t([L^\dagger,X])\,{\rm d} A_t\,,
\end{equation}
where $\mathcal{L}$ is the GKSL dual generator in the Heisenberg picture:
$$
\mathcal{L}(X)=i\,[H,X]\,+\,
L^\dagger\, X\,L-\frac{1}{2}\Big\{L^\dagger\, L\,,\,X\Big\}\ .
$$
From the quantum Langevin equation one retrieves the GKSL master equation as the corresponding quantum Fokker-Planck equation:
choose $\rho_S\otimes\ket{\Omega}\bra{\Omega}$ as initial system+environment state, with $\rho_S$ a generic open system state and $\ket{\Omega}$ the environment vacuum.
Then, from
\begin{equation}
    \label{HP-FP}
\Tr_S\Big(\rho_S(t)\,X\Big):=\Tr_{S+\mathcal{F}}\Big(\rho_S\otimes \ket{\Omega}\bra{\Omega}\,j_t(X)\Big)
\end{equation}
and the fact that $A_t\ket{\Omega}=0$, one finds  
$$
{\rm d}\rho_S(t)=\Big(-i\,[H,\rho_S(t)]\,+\,
L\, \rho_S(t)\,L^\dag-\frac{1}{2}\Big\{L^\dagger\, L\,,\,\rho_S(t)\Big\}\Big)\,{\rm d}t\ .
$$
Despite the similarities between~\eqref{HP-FP} and~\eqref{red-dyn},  the operators $U_t$  in the latter do not compose as a group, rather as a cocycle; namely,
$$
U_{s+t}=\Theta_s(U_t)U_s\ ,
$$
where (see~\eqref{noise-op})
$$
\Theta_s(A_t)=A_{[s,t+s]}=A_{t+s}-A_s
$$
implements the forward time-shift on the environment Fock space. The meaning of the cocycle property is that, in the splitting $[0,t+s]=[0,s]\cup[s,t+s]$, {while the operator $U_s$ depends on the noise from the interval $[0,s]$, in order to match it, 
the operator $U_t$ has to be adapted to the noise from $[s,s+t]$ and thus shifted by $\Theta_s$} in agreement with Remark~\ref{rem:adapt}.

\begin{Remark}
\label{rem:Accardi}
In the approach known as \textit{Quantum Stochastic Limit} developed by Accardi, Lu and Volovich~\cite{ALV2002}, 
the quantum stochastic tools of Hudson and Parthasarathy are recovered within the broader perspective of the Davies weak-coupling limit.
The latter typical long-time scaling is considered by taking into account the whole system-environment interaction, that is  both system and environment together, without tracing away the latter.    
\end{Remark}

\subsection{Ellis-Hagelin-Nanopoulos-{S}rednicki and Banks-Susskind-Peskin analysis (1984)}

Surprisingly, knowledge of the canonical GKLS result was not yet common across
theoretical physics in 1984. {In~\cite{EHNS84} Ellis, Hagelin, Nanopoulos and Srednicki 
argued that gravitational effects at the Planck's scale could spoil unitarity and cause decoherence.  
They wrote down a master equation for the neutral $K$ mesons, treated as a two-level system, involving  three new phenomenological parameters, chosen to ensure positivity of the resulting dissipative dynamics.
In a more general context, their proposal was then criticized by Banks, Susskind, and Peskin in~\cite{Banks1984} who pointed to the violation of locality and energy-momentum conservation.
Both groups were evidently unaware of the 1976 GKS and Lindblad papers and drew from}
Hawking's proposal that a
pure state might evolve fundamentally into a mixed state.  Their papers cites
neither earlier work, and complete positivity plays no role in their argument.
They imposed ordinary positivity of the density matrix and explicitly stated
that they did not know necessary conditions for preserving it. 

For a finite-dimensional Hilbert space, let
$\{Q^\alpha\}_{\alpha=0}^{N^2-1}$ be an orthogonal Hermitian operator basis,
with $Q^0=\oper$.  Hermiticity and trace preservation put a general
time-local linear equation in the form

\begin{equation}
 \dot\rho_t =-i[H_0,\rho_t]
 -\frac12\sum_{\alpha,\beta\ne0}h_{\alpha\beta}
 \left(Q^\beta Q^\alpha\rho_t +\rho_t \, Q^\beta Q^\alpha
       -2Q^\alpha\rho_t \, Q^\beta\right),
 \qquad h_{\alpha\beta}=h_{\beta\alpha}^*.
 \label{eq:bsp-generator}
\end{equation}
Interestingly, and apparently without knowledge of Bausch's 1966 derivation \cite{Bausch1966} or the subsequent 
Kossakowski results on semigroups of positive trace-preserving maps \cite{AK-1,AK-2}, Banks, Susskind, and Peskin independently recovered essentially the canonical master-equation structure. Now, the key problem is to implement the requirement that $\rho$ remains positive in the course of time. The authors of \cite{Banks1984} honestly stated ``{\em we do not know what conditions are necessary to insure these properties, but we can state some simple sufficient conditions}.'' Clearly, $\rho$ remains positive if  the Kossakowski matrix $h_{\alpha\beta}$ is positive definite. Hence, they obtained the GKLS form without invoking complete positivity.

{Another important observation 
they made is} that when $h_{\alpha\beta}$ is not only Hermitian but real symmetric, then 
the von Neumann entropy $S(\rho)=-\operatorname{Tr}(\rho\log\rho)$ satisfies

\begin{equation}
    \frac{d}{dt}S(\rho_t) \geq0 .
\end{equation}
This problem was further analyzed in \cite{Benatti_1988} (cf. also \cite{Aniello}). Moreover, when all jump operators $Q^\alpha$ are Hermitian they gave the dissipator a useful stochastic interpretation.  Ordinary unitary dynamics driven by a classical white-noise
Hamiltonian

\begin{equation}
 H(t)=H_0+\sum_\lambda j_\lambda(t)\, Q^\lambda,
 \qquad
 \mathbb{E}[j_\lambda(t)]=0,
 \qquad
 \mathbb{E}\!\left[j_\lambda(t)j_\mu(t')\right]
 = h_{\lambda\mu}\delta(t-t') ,
 \label{eq:bsp-random-source}
\end{equation}
reproduces $\rho_t$ after averaging over the sources.  The loss
of coherence is therefore the ensemble effect of uncontrolled classical
driving.  In an individual realization the source is time dependent and can
add or remove energy, already exposing the first difficulty with treating the
averaged equation as a fundamental law for a closed system.

 The main physical question addressed in \cite{Banks1984} was whether such a linear, local-in-time law can also satisfy the usual requirements of local quantum field theory.  The authors argued that a
genuinely coherence-destroying term leads either to energy--momentum
non-conservation or to spatial nonlocality. A translation-covariant extension of \eqref{eq:bsp-generator} has the form

\begin{equation}
 \dot\rho_t =-i\left[\int d^3x\,\mathcal{H}(\mathbf{x}),\rho\right]
 -\frac12\int d^3x\,d^3y\,
 h_{\alpha\beta}(\mathbf{x}-\mathbf{y})\\
 \left(
 \{Q^\beta(\mathbf{y})Q^\alpha(\mathbf{x}),\rho_t\}
 -2Q^\alpha(\mathbf{x})\rho_t Q^\beta(\mathbf{y})\right),
 \label{eq:bsp-field-generator}
\end{equation}
where
$h_{\alpha\beta}(\mathbf{x}-\mathbf{y})
 =h_{\beta\alpha}^*(\mathbf{y}-\mathbf{x})$.  Its dissipative part is, in
momentum space,

\begin{equation}
 \mathcal{D}(\rho_t)=-\frac12\int\frac{d^3p}{(2\pi)^3}\,
 \widetilde h_{\alpha\beta}(\mathbf{p})
 \left(\{ \widetilde{Q}^{\beta\dagger}(-\mathbf{p})\widetilde{Q}^\alpha(\mathbf{p}),\rho_t\}
 -2\widetilde{Q}^\alpha(\mathbf{p})\rho_t \widetilde{Q}^{\beta\dagger}(-\mathbf{p})\right).
 \label{eq:bsp-fourier-generator}
\end{equation}
It is shown \cite{Banks1984} that  nontrivial, local-in-time pure-to-mixed modification
constructed from local field operators cannot simultaneously retain ordinary spatial locality and
exact energy–momentum conservation. A short-range kernel produces energy and momentum,
whereas restricting the kernel to conserved or zero-momentum modes makes the evolution spatially
nonlocal and, in the constant-kernel limit, can violate cluster decomposition. This problem was further analyzed in \cite{Unruh_1995} (see also the recent review \cite{Unruh_2017}).

{
\begin{Remark}
    \label{rem:BSP}
The idea of~\cite{EHNS84} was taken up by Huet and Peskin in~\cite{HuetPeskin95} who proposed to use the rich neutral Kaon phenomenology to estimate the new phenomenological parameters appearing in the Kaon master equation proposed in~\cite{EHNS84}. In~\cite{BF96,BF97}, Benatti and Floreanini imposed that the phenomenological 
parameters would ensure complete positivity and proposed to use the K meson phenomenology to check whether the CP-constraints coming form the GKSL structure of the generator are satisfied or not.  
In other words, the idea was to use the abundance and accuracy of meson data to establish whether a phenomenological description of Planck's scale and quantum gravity effects by means of a GKSL reduced dynamics is feasible or not.
That line of ideas was then extended to correlated mesons~\cite{BF98,BFR2001} where dissipation and complete 
positivity were put face-to-face with quantum correlations and entanglement.
\\
Decoherence in particle physics with reference to neutral mesons phenomenology, in particular in connection to entanglement and Bell's inequalities has been addressed in~\cite{Hiesmayr2001,BGH2003,H2017}, while
in~\cite{FFP2021,BFFGM2024} the idea is developed to use hadron collider observables to expose violations of Bell's inequalities in higher dimensional settings  using the spin and polarization correlations of particles produced 
at high energies.
\end{Remark}
}

\subsection{Dalibard--Castin--M{\o}lmer: quantum-jump unraveling (1992)}
\label{sec:quantum-jump-unraveling}

The GKLS master equation gives the deterministic evolution of the density
operator, but it does not by itself display the individual random events that
may occur when the environment is continuously monitored.  In 1992, Dalibard,
Castin, and M{\o}lmer proposed the Monte Carlo wave-function (MCWF) method for
dissipative processes in quantum optics
\cite{DalibardCastinMolmer1992}.  A closely related construction was developed
independently by Dum, Zoller, and Ritsch \cite{DumZollerRitsch1992}; see also
the review \cite{PlenioKnight1998} and \cite{Carmichael1993}.  The central idea is to replace the master
equation for an ensemble by a stochastic evolution of normalized wave
functions.  Between photon detections the state evolves with a non-Hermitian
effective Hamiltonian, whereas a detection produces a discontinuous quantum
jump.  Averaging the pure-state projectors over all detection records exactly
recovers the density operator.

Consider a GKLS generator written as
\begin{equation}
  \frac{d\rho_t}{dt}=\mathcal{L}(\rho_t)
  =-i[H,\rho_t]
   +\sum_{\mu=1}^{m}\left(
      L_\mu\rho_t L_\mu^\dagger
      -\frac{1}{2}\{L_\mu^\dagger L_\mu,\rho_t\}
    \right) ,
  \label{eq:unraveling-gkls}
\end{equation}
and let us introduce

\begin{equation}
  R:=\sum_{\mu=1}^{m}L_\mu^\dagger L_\mu,
  \qquad
  H_{\mathrm{eff}}:=H-\frac{i}{2}R .
  \label{eq:effective-Hamiltonian}
\end{equation}
For a normalized state vector $|\psi\rangle$, define the channel intensities
\begin{equation}
  r_\mu(\psi):=
  \langle\psi|L_\mu^\dagger L_\mu|\psi\rangle,
  \qquad
  r(\psi):=\sum_\mu r_\mu(\psi).
  \label{eq:jump-intensities}
\end{equation}
During an infinitesimal interval $[t,t+dt]$, the state $|\psi_t\rangle$ at time $t$ changes according to
the following alternatives:

\begin{enumerate}

\item if no jump is detected, which occurs with probability
$dp_0=1-r(\psi_t)dt$, then
\begin{equation}
  |\psi_t\rangle\longmapsto
  |\psi_t^{(0)}\rangle
  :=\frac{(1-iH_{\mathrm{eff}}dt)|\psi_t\rangle}
          {\sqrt{1-r(\psi_t)dt}}.
  \label{eq:no-jump-update}
\end{equation}

    \item If a jump is detected in channel $\mu$, which occurs with probability
\begin{equation}
  dp_\mu=r_\mu(\psi_t)\,dt,
  \label{eq:jump-probability}
\end{equation}
then
\begin{equation}
  |\psi_t\rangle\longmapsto
  |\psi_t^{(\mu)}\rangle
  :=\frac{L_\mu|\psi_t\rangle}
          {\sqrt{r_\mu(\psi_t)}}.
  \label{eq:jump-update}
\end{equation}

\end{enumerate}

Indeed,
\begin{equation}
  \bigl\|(1-iH_{\mathrm{eff}}dt)|\psi_t\rangle\bigr\|^2
  =1-r(\psi_t)dt+O(dt^2),
\end{equation}
so the loss of norm under the effective evolution is precisely the total
probability of a jump.
Equivalently, the normalized trajectory satisfies the nonlinear stochastic
Schr{\"o}dinger equation
\begin{align}
  d|\psi_t\rangle={}
  \left[-iH-\frac{1}{2}\sum_\mu
    \bigl(L_\mu^\dagger L_\mu-r_\mu(\psi_t)\,1\!\mathrm{l}\bigr)
  \right]|\psi_t\rangle\,dt \notag  +\sum_\mu\left(
    \frac{L_\mu}{\sqrt{r_\mu(\psi_t)}}-1\!\mathrm{l}
  \right)|\psi_t\rangle\,dN_\mu(t),
  \label{eq:jump-sse}
\end{align}
where $dN_\mu(t)\in\{0,1\}$ are Poisson increments satisfying, conditionally
on the state at time $t$,
\begin{equation}
  \mathbb{E}\!\left[dN_\mu(t)\,\middle|\,\psi_t\right]
  =r_\mu(\psi_t)dt,
  \qquad
  dN_\mu(t)dN_\nu(t)=\delta_{\mu\nu}dN_\mu(t).
  \label{eq:poisson-increments}
\end{equation}
Let $P_t=|\psi_t\rangle\langle\psi_t|$ be the corresponding rank-1 projector.  The equivalence with the GKLS
equation follows directly by averaging one infinitesimal step.  Conditional
on $|\psi_t\rangle$, Eqs.~\eqref{eq:jump-update} and
\eqref{eq:no-jump-update} give
\begin{align}
  \mathbb{E}[P_{t+dt}\mid\psi_t]
  ={}&(1-rdt)\rho_t^{(0)}
      +\sum_\mu r_\mu dt\,P_t^{(\mu)} \notag\\
  ={}&(1-iH_{\mathrm{eff}}dt)P_t
      (1+iH_{\mathrm{eff}}^\dagger dt)
      +dt\sum_\mu L_\mu P_tL_\mu^\dagger+O(dt^2) \notag\\
  ={}&P_t+dt\,\mathcal{L}(P_t)+O(dt^2).
  \label{eq:conditional-projector-average}
\end{align}
Consequently, if
\begin{equation}
  \rho_t=\mathbb{E}\bigl[|\psi_t\rangle\langle\psi_t|\bigr],
  \label{eq:ensemble-unraveling}
\end{equation}
then linearity of $\mathcal{L}$ yields
the GKLS master equation $\dot\rho_t=\mathcal{L}(\rho_t)$.  A mixed initial state causes no difficulty:
one first chooses any ensemble
$\rho_0=\sum_k p_k|\psi_k\rangle\langle\psi_k|$ and samples the initial wave
function with probabilities $p_k$.  Thus the stochastic process
\emph{unravels} the deterministic master equation.

It is essential that an unraveling is not unique.  For example, a unitary
mixing of the jump operators,
\begin{equation}
  L'_\alpha=\sum_\mu U_{\alpha\mu}L_\mu,
  \qquad U^\dagger U= \oper,
  \label{eq:unitary-jump-mixing}
\end{equation}
leaves the dissipator in Eq.~\eqref{eq:unraveling-gkls} unchanged but generally
changes the individual jumps.  More generally, the transformation
\begin{equation}
  L'_\mu=L_\mu+c_\mu 1\!\mathrm{l},
  \qquad
  H'=H+\frac{1}{2i}\sum_\mu
  \left(c_\mu^*L_\mu-c_\mu L_\mu^\dagger\right)
  \label{eq:affine-jump-gauge}
\end{equation}
also represents the same GKLS generator.  Physically, different unravelings
correspond to different measurements on the output field.  Direct photon
counting produces jump trajectories, while homodyne or heterodyne detection
produces continuous diffusive trajectories.  Already in
Ref.~\cite{DalibardCastinMolmer1992}, measuring different components of the
angular momentum of the emitted photon led to different stochastic pictures
of population trapping, although all choices reproduced the same optical
Bloch equation after averaging.

\begin{Remark}
    The unraveling can be substantially more efficient numerically than a direct
integration of the master equation.  For a Hilbert space of dimension $N$, a
wave function contains $N$ complex amplitudes, whereas the density operator
contains $N^2$ matrix elements.   This reduction from an
$N^2$-dimensional deterministic problem to repeated $N$-dimensional pure-state
problems was one of the principal motivations for the MCWF method in
Ref.~\cite{DalibardCastinMolmer1992}.
\end{Remark}

\begin{Remark} There is an extensive body of literature on stochastic unravelings.  Belavkin developed the general theory of quantum nondemolition filtering and posterior quantum dynamics, beginning with his work on
optimal quantum filtration in the late 1970s. He derived stochastic
equations for the state conditioned on a continuous measurement record,
including both diffusive observations and photon-counting processes
\cite{Belavkin1989Wave,Belavkin1989Counting,Belavkin1992}. Di{\'o}si formulated continuous quantum measurement in   It{\^o} calculus and derived a diffusive stochastic evolution for the state
  vector \cite{Diosi1986, Diosi1988}.  This work provided an early direct bridge between
  continuous monitoring and stochastic Schr{\"o}dinger equations.
Gisin and Percival introduced quantum-state diffusion as a continuous,
  norm-preserving unraveling of Markovian master equations
  \cite{GisinPercival1992}.  Wiseman and Milburn clarified the measurement
  interpretation of both jump and diffusion processes
  \cite{WisemanMilburn1993} and Carmichael developed the photodetection-based
  quantum-trajectory theory systematically in  \cite{Carmichael1993}.
Di{\'o}si, Gisin, and Strunz extended quantum-state diffusion to
  non-Markovian reservoirs \cite{DiosiGisinStrunz1998, StrunzDiosiGisin1999}.
Piilo, Maniscalco, H{\"a}rk{\"o}nen, and Suominen introduced the
  non-Markovian quantum-jump method \cite{PiiloEtAl2008, PiiloEtAl2009}. See also \cite{Nori2012} for a numerical framework to simulate open system dynamics, including GKLS master equations, Bloch-Redfield theory,  and Monte Carlo quantum trajectories.

\end{Remark}

\begin{Remark}
A complementary microscopic picture of unraveling is provided by
collision models, also known as repeated-interaction models.
The system interacts successively with fresh, initially uncorrelated
ancillas prepared in the same state. Tracing out each outgoing
ancilla yields a discrete reduced evolution which, under an
appropriate scaling of the interaction strength and collision
duration, converges to a GKLS semigroup. Collision models thus provide a concrete connection between
microscopic interactions, environmental monitoring, and the
stochastic description of open quantum dynamics. This approach was initiated by Rau \cite{Rau1963} and popularized by Scarani et. al. \cite{Scarani2002} in quantum information science to model homogenization, qubit thermalization, and quantum channels (see \cite{Ciccarello2022} for a comprehensive review).

\end{Remark}

\subsection{Pechukas: challenging complete positivity (1994)}

In all preceding sections, a somewhat tacit assumption was taken: Alicki in~\cite{AvsP95} calls it a \textit{fundamental positivity condition}. Namely, the fact that only complete positivity can give physical legitimacy to linear dynamical maps $\Lambda_t$ describing the time-evolution of the states of an open quantum system $S$. 

Such a strong stance has two motivations; one is  fairly general and abstract. Indeed, a non-completely positive (NCP) $\Lambda_t$ is such that $\Lambda_t\otimes{\rm id}_k$ fails to preserve the positivity of 
some entangled state of the compound system $S+S_k$, for some $k$-level system $S_k$.

\begin{Remark}
\label{rem:AvP}
If $S$ is also a finite, $d$-level system, then the Choi-Jamiołkowski isomorphism ensures that $\Lambda_t\otimes{\rm id}_d$ acting on the completely symmetric $d$-dimensional , projector $P^+$ has negative eigenvalues appearing in the spectrum of $\Lambda_t\otimes{\rm id}_d[P^+]$ which cannot then be a legitimate quantum state. In other words, NCP maps tensorized with the identity map on generic, finite level systems $S_k$ cannot provide physically acceptable transformations for all entangled 
states of the coupled systems $S+S_k$ as they may generate negative probabilities acting on some of them.
\end{Remark}

The second motivation is related to the derivation of an open system reduced dynamics $\Lambda_t$ as done in the Section~\ref{sec:QR} 
(see Eq.~\eqref{red-dyn}). In that case, complete positivity is a necessary consequence of the fact that both the unitary dynamics $\mathcal{U}_t$ and the partial trace $\Tr_R$ over the environment are CP maps. 

In~\cite{Pechukas}, Pechukas argued against the necessity of complete positivity and advocated the physical legitimacy of NCP (actually, also of non-positive) maps, provided a restriction of their initial states is considered.
Pechukas challenge was taken on by Alicki in~\cite{AvsP95} who showed that renouncing complete positivity is not strictly necessary. Pechukas replied in~\cite{Pechukas94} that according to him the price to pay for salvaging complete positivity was physically higher than dispensing with it. Such an inspiring short exchange of radically different points of view ignited a debate which is still going on, to which, quite interestingly, important contributions came from Sudarshan and collaborators~\cite{Jordan2004,Shaji2005}. Asking \textit{who is afraid of non completely positive maps?}, they seemingly go back to the early stages of open quantum systems theory where, as summarized in Section~\ref{S:Sudarshan}, the notion of complete positivity and its implications were not yet fully appreciated.

In~\cite{Pechukas} Pechukas criticized the general argument in favor of complete positivity by observing that when, as in Eq.~\eqref{wcl1} or Eq.~\eqref{eq:davies-scaling}
$\Lambda_t$ is a reduced dynamics derived from the coupling of $S$ to a reservoir $R$, the intervention of the system $S_k$ is rather 
artificial as it does not interact either with $R$ or $S$ and is inert, only being possibly entangled with $S$. Yet, physical constraints on the diagonal and off-diagonal decay times as those in~\eqref{hierarchy} emerge exactly because of the specific GKSL form of the generator dictated by complete positivity.  This fact, in Pechukas own words (adapted to the notation of the present discussion) \textit{is very
powerful magic: $S_k$ sits apart from $S + R$ and does
absolutely nothing; by doing so, it forces the motion
of $S$ to be completely positive, with dramatic physical
consequences, such as in~\eqref{hierarchy} for exponential two-state
relaxation.}

As for the second motivation, the crucial issue in Eq.~\eqref{red-dyn} is that, in order to derive a reduced dynamics for the open system states $\rho_S$, one has first 
to embed them into the space of states $\rho_{SR}$ of the compound system $S+R$. In~\eqref{red-dyn}, the assignment map $\Phi$, as Pechukas calls it, is such that $\rho_S\mapsto \Phi[\rho_S]=\rho_S\otimes\rho_R$.
It is thus consistent with the absence of initial correlations between system and reservoir, which is far from being the only possible scenario.

As pointed out by Alicki in his comment~\cite{AvsP95} on Pechukas argument, the following three properties look reasonable constraints to impose on any assignment map: 
\begin{itemize}
\item 
\textbf{Convex Linearity}: for all $0\leq\lambda\leq 1$ and for all states $\rho_{1S,2S}$ of $S$,
$$\Phi[\lambda\rho_{1S}+(1-\lambda)\rho_{2S}]=\lambda\Phi[\rho_{1S}]+
(1-\lambda)\Phi[\rho_{2S}]\ ;
$$ 
\item
\textbf{Consistency}: for all states $\rho_S$ of $S$,
$$
{\rm Tr}_R\Phi[\rho_S]=\rho_S\ ;
$$
\item 
\textbf{Positivity}: $\Phi[\rho_S]\geq 0$ for all states $\rho_S$ of $S$.
\end{itemize}
The first condition ensures that the convex structure of the set of quantum states is preserved by the embedding , the second one that the embedding retrieves the given initial system states when the reservoir is eliminated and, finally, the third one complies with the embedded system states being states of the compound system. Then, one has~\cite{Pechukas94},

\begin{Theorem}
\label{Th_Pechukas}
Assignment maps fulfilling the above conditions must be of the form 
$\Phi[\rho_S]=\rho_S\otimes\rho_R$ with a same reservoir state $\rho_R$ for all system states $\rho_S$.
\end{Theorem}

In~\cite{Pechukas94}, the proof is provided for $S$ a $2$-level system and $R$ a reservoir described by a Hilbert space with a discrete  orthonormal basis $\{\ket{r}\}$. It consists in two steps: firstly, one shows that that the assignment map is of the factor form for all system pure states $\ket{0}\bra{0}$  with an associated $\ket{0}$-dependent reservoir state.
Indeed, consider a pure state $\rho_S=\ket{0}\bra{0}$ of $S$.
The consistency and positivity conditions yield:
$$
\rho_S={\rm Tr}_R\Phi[\rho_S]=\sum_{r'}\bra{r'}\rho_{RS}\ket{r'}\geq \bra{r}\rho_{RS}\ket{r}\ ,\quad \forall r\ .
$$
If $\ket{1}$ is orthogonal to $\ket{0}$, then, again by positivity, 
$$
0=\bra{1}\rho_S\ket{1}\geq \bra{1,r}\rho_{RS}\ket{1,r}\Longrightarrow 
\sqrt{\rho_{RS}}\ket{1,r}=0\quad \forall r\ .
$$
It thus follows that
$$
\rho_{SR}=\sum_{s,s'=0,1}\sum_{r,r'}\bra{s',r'}\rho_{SR}\ket{s,r}\,\ket{s'}\bra{s}\otimes\ket{r'}\bra{r}
=\ket{0}\bra{0}\otimes
\sum_{r,r'}\bra{0,r'}\rho_{RS}\ket{0,r}\,\ket{r'}\bra{r}\ ,
$$
with an $\ket{0}$-dependent reservoir state $\rho_R=\sum_{r,r'}\bra{0,r'}\rho_{RS}\ket{0,r}\,\ket{r'}\bra{r}$.

Secondly, by using the convexity condition, one eliminates the dependence of the reservoir state on the system state.
Indeed, consider two convex orthogonal decompositions of the totally depolarized state which, by a suitable unitary rotation, can always be chosen to be
$$
\frac{1}{2}\mathds{1}=\frac{1}{2}\begin{pmatrix}1&0\cr0&0\end{pmatrix}+\frac{1}{2}\begin{pmatrix}0&0\cr0&1\end{pmatrix}=
\frac{1}{2}\begin{pmatrix}|\alpha|^2&\alpha\beta^*\cr\alpha^*\beta&|\beta|^2\end{pmatrix}+\frac{1}{2}\begin{pmatrix}|\beta|^2&-\alpha\beta^*\cr-\alpha^*\beta&|\alpha|^2\end{pmatrix}\ ,
$$
with $\alpha,\beta\in\mathbb{C}$ such that $|\alpha|^2+|\beta|^2=1$. The previous result then gives
\begin{eqnarray*}
\Phi[\mathds{1}]=\begin{pmatrix}1&0\cr0&0\end{pmatrix}\otimes\rho_++\begin{pmatrix}0&0\cr0&1\end{pmatrix}\otimes\rho_-&=&
\begin{pmatrix}\rho_+&0\cr0&\rho_-\end{pmatrix}\\
&=&\begin{pmatrix}|\alpha|^2&\alpha\beta^*\cr\alpha^*\beta&|\beta|^2\end{pmatrix}\otimes\sigma_++
\begin{pmatrix}|\beta|^2&-\alpha\beta^*\cr-\alpha^*\beta&|\alpha|^2\end{pmatrix}\otimes\sigma_-\\
&=&\begin{pmatrix}|\alpha|^2\sigma_++|\beta|^2\sigma_-&\alpha\beta^*(\sigma_+-\sigma_-)\\
\alpha^*\beta(\sigma_+-\sigma_-)&|\beta|^2\sigma_++|\alpha|^2\sigma_-
\end{pmatrix}\ ,
\end{eqnarray*}
where $\rho_{\pm}$, respectively $\sigma_{\pm}$ are reservoir states that may depend on the system states projecting onto the orthogonal state vectors $(1,0)^T$,
$(0,1)^T$, respectively $(\alpha,\beta)^T$, $(\beta^*,-\alpha^*)^T$. Equating the matrix entries appearing in the previous equality, one gets $\sigma_+=\sigma_-=\rho_+=\rho_-$.

Therefore, any assignment map satisfying convexity, consistency and positivity cannot but lead to a CP reduced dynamics.
On the other hand, if system and reservoir strongly interact, the preparation of the joint system-reservoir initial state 
can hardly achieve a factorized one, so that the assignment map need violate some of the natural conditions imposed on it.

While convexity is hard to renounce as it would give rise to a scenario far away from the linear framework of Quantum Mechanics, as regards the other two, Alicki's stance {is: better violate consistency rather than positivity, whereas Pechukas~\cite{Pechukas94} is ready to renounce positivity and thus complete positivity.}

Sticking to this latter choice, what about the appearance of negative probabilities, for instance in the spectrum of
$\Lambda_t\otimes{\rm id}_d[P^+]$, with $P^+$ the symmetric projection for $S_d+S_d$, when $\Lambda_t$ is NCP?

Certainly, if $\Lambda_t$ is at least positive, separable states of the form $\rho_{sep}=\sum_\alpha\lambda_\alpha \rho_\alpha\otimes\rho^{(k)}_\alpha$, with $\lambda_\alpha\geq 0$, $\sum_\alpha\lambda_\alpha=1$ and $\rho_\alpha$ and $\rho_{\alpha}^{(k)}$, states of $S$ respectively $S_k$, are mapped into states by $\Lambda_t\otimes{\rm id}_k$ even for an NCP $\Lambda_t$, and not all entangled states of $S+S_k$ assume negative eigenvalues.
One could then salvage NCP maps as physically legitimate by restricting the action of $\Lambda_t\otimes{\rm id}_k$ to subsets of initial states of $S+S_k$ for all $k$, namely 
by taking as legitimate initial conditions only those states that remain positive under the action of $\Lambda_t\otimes{\rm id}_k$.

Such an \textit{ad hoc} selection of initial conditions, a so-called \textit{compatibility domain}, is the conclusive recommendation of Pechukas, a point of view taken up later on 
in~\cite{Shaji2005,Jordan2004,Rodriguez2008,Cuffaro2013}. 

Evidently, finding the compatible domain of an NCP map $\Lambda_t$ asks for the mathematical characterization, for all $k$, of all entangled states of $S+S_k$ that are mapped into states by $\Lambda_t\otimes{\rm id}_k$. Further, the restriction to them of the action of $\Lambda_t\otimes{\rm id}_k$ demands a physical selection mechanism. For instance, in the approach known as \textit{slippage of initial conditions}~\cite{Suarez92,Gnutzmann96,GaspardNagaoka1999,Whitney2008} where it is assumed to occur during a transient interval of time before the NCP dynamics sets in. However,  explicit and formal implementations of the selection of harmless initial conditions do not seem to be available, yet.
Finally, since entanglement becomes typical with increasing dimension, one suspects~\cite{FDS-2026} that, for $\Lambda_t$ NCP when $d\gg1$, the compatibility domains for $\Lambda_t\otimes{\rm id}_k$ on $S_d+S_d$ may become thinner and thinner.

\section{Conclusions}

In 2017, the numbers 3 and 4 of Volume 24 of \textit{Open Systems \& Information Dynamics}
celebrated the fortieth anniversary of the GKLS theorem. In the opening contribution, the authors
explored what they called, in the Conclusions, ``a number of concomitant factors that led to the
precise mathematical formulation of the evolution equations for open quantum dynamical systems.''
The present contribution, written on the occasion of the first half-century of the GKLS theorem,
returns to that history with broader scope and deeper retrospection.

The narrative we have traced begins well before 1976. The 1961 paper of Sudarshan, Mathews
and Rau already contained the operator-sum representation for completely positive maps, though
its authors did not yet distinguish complete positivity from mere positivity. Bausch in 1966
independently derived the correct infinitesimal architecture of the generator, and Belavin,
Zel'dovich, Perelomov and Popov in 1969 identified positivity of the coefficient matrix as the key
condition ensuring preservation of the density matrix. Kossakowski's semigroup analysis of
1972--1973 provided the geometric and functional-analytic scaffolding, and Davies's weak-coupling
limit of 1974 supplied the first rigorous microscopic derivation. Yet, despite all these
anticipations, one decisive conceptual step remained to be made.

That step was the recognition of \emph{complete positivity} as the correct and physically
indispensable requirement for the evolution of an open quantum system. This is the true
intellectual achievement of Gorini, Kossakowski and Sudarshan, and independently of Lindblad,
in 1976. Earlier authors had worked with positivity, block positivity, conditional positivity, and
various sufficient conditions --- each capturing part of the truth. What GKS and Lindblad
understood, and made mathematically precise, is that the evolution of an open system must remain
physically consistent when that system is coupled, even without any interaction, to an arbitrary
ancilla. This requirement --- that the extended dynamics on system plus ancilla must preserve
positivity for all states, including entangled ones --- is precisely complete positivity. It is not
merely a convenient mathematical condition: it is the condition that Nature imposes.

In this sense, Physics itself has chosen the mathematical path of complete positivity. The
GKLS structure theorem is not simply one among several equivalent formulations of Markovian
open dynamics; it is the one formulation that is stable under the most general physical scenario,
that is consistent with the possibility of entanglement, and that survives every test that quantum
information theory, quantum thermodynamics, and quantum optics have subsequently brought to
bear upon it. The dramatic and sustained growth of citations to both the GKS and Lindblad papers
--- illustrated in Figure~1 and reaching its highest values precisely around the fiftieth anniversary
--- is a quantitative reflection of this fact.

The question of why complete positivity, rather than mere positivity, is the correct condition
naturally invites reflection on how one compares different but related developments. Here the
words of Noam Chomsky offer an unexpectedly apt methodological principle~\cite{Chomsky1985}:
\begin{quote}
\textit{If you take any two historical events and you ask whether there are similarities and
differences, the answer is always going to be both ``yes'' and ``no.'' At some sufficiently fine
level of detail there will be differences, and at some sufficiently abstract level there will be
similarities. The question we want to ask in the two cases we are considering, [\ldots] is whether
the level at which there are similarities is, in fact, a significant one.}
\end{quote}
This principle, formulated by Chomsky in a political and historical context, applies with
remarkable force to the history of open quantum dynamics. The works of Sudarshan \textit{et
al.}, Bausch, Belavin \textit{et al.}, Kossakowski, and Davies are all similar to the GKLS theorem
at some level: they involve density matrices, semigroups, positivity, and dissipative generators.
But the \emph{significant} level of similarity --- the one that matters physically and
mathematically --- is the level of complete positivity, and it is precisely at that level that the
1976 papers stand apart from everything that preceded them. The similarities with earlier works
are real, and we have honoured them throughout this review; but the difference introduced by
complete positivity is the significant one.

The subsequent developments examined in Sections~10--16 confirm this assessment. Alicki's
quantum detailed balance, the Spohn--Frigerio theory of relaxation and approach to equilibrium,
Spohn's entropy production inequality, Lindblad's analysis of quantum regression and
Markovianity, the Hudson--Parthasarathy quantum stochastic calculus, the quantum-jump
unraveling of Dalibard, Castin and M{\o}lmer, and the Pechukas debate on complete positivity ---
all of these developments grow naturally and inevitably from the GKLS structure. They are not
merely applications; they are consequences of the fact that complete positivity is the right
framework.

The story is not over. Unbounded generators, quantum thermodynamics beyond weak coupling,
the characterisation of stationary states in infinite-dimensional systems, non-Markovianity and
information backflow, and the foundations of quantum stochastic processes all remain active and
partially open. The hope, supported by the contributions to this special volume of \textit{Open
Systems \& Information Dynamics}, is that the next anniversary will see some of these questions
answered --- and that complete positivity will continue to be recognised, as it deserves, as the
cornerstone on which the entire edifice rests.

\section*{Acknowledgements}

D.C. was supported by the Polish National Science
Centre project No. 2024/55/B/ST2/01781.
SP acknowledges financial support from Projects PN-RIC Q-SUD (Q-SUD B99H26000380007) and PN-RIC QUANTAS (SYNERGIA B99H26000410007).

\end{document}